\documentclass[twocolumn,epjc3]{svjour3}          
\journalname{Eur. Phys. J. A}
\usepackage{cuted}
\usepackage{slashed}
\def\nn{\nonumber\\}        
\usepackage{graphicx}
\usepackage{amsbsy}
\input{epsf}
\def\sls {\sqrt{\lambda_S}}
\def\slx {\sqrt{\lambda_X}}
\begin{document}

\title{ 
Charge-asymmetric correlations in elastic lepton- and antilepton-proton scattering from real photon emission
}
\author{ A. Afanasev\thanksref{e1,addr1}
            \and
     A. Ilyichev\thanksref{e2,addr2,addr3}}

\thankstext{e1}{e-mail: afanas@gwu.edu}
\thankstext{e2}{e-mail: ily@hep.by}

\institute{Department of Physics,
The George Washington University,
Washington, DC 20052 USA\label{addr1}
\and
 Belarusian State University,
220030  Minsk,  Belarus\label{addr2}
\and
Institute for Nuclear Problems,
 Belarusian State University,
220006  Minsk,  Belarus\label{addr3}
}

\date{Received: date / Accepted: date}

\maketitle
\begin{abstract}
Observation of charge asymmetry by comparing electron and positron, or muon and anti-muon, scattering on a hadronic target presently serves as an experimental tool to study two-photon exchange effects. In addition to two-photon exchange, real photon emission also contributes to the charge asymmetry.
We present a theoretical formalism, explicit expressions, and a numerical analysis of hard photon emission for the charge asymmetry in lepton- and antilepton-proton scattering. Different kinematic conditions are considered, namely, either fixed transferred momentum squared or a fixed lepton scattering angle. 
The infrared divergence from real photon emission is treated by the Bardin-Shumeiko technique and canceled with the soft part of the two-photon exchange contribution extracted and calculated using Tsai approach.  All final expressions are obtained beyond the ultrarelativistic approximation with respect to the lepton mass that allows to evaluate numerically of the considered effects not only for ultrarelativistic leptons (JLab) and but for moderately relativistic (MUSE) kinematics, too. 
\end{abstract}

\section{Introduction}
The most precise measurements allowing us to study a quark-gluon structure of hadrons come from experiments on lepton-proton scattering, when a structureless pointlike particle -- a lepton -- interacts with the simplest hadronic structure -- a proton. A rather small value of the fine structure constant, $\alpha \sim 1/137 $,   makes the electromagnetic interaction amenable to a perturbative treatment in the context of quantum field theory.

The fundamental observables characterizing the internal structure of the proton are its electric and magnetic form factors. However, the measurements of their ratio in unpolarized and polarized  electron elastic scattering data differ up to three  times at $Q^2\approx $ 6 GeV$^2$ \cite{Andivahis94,Qattan05,Jones,Gayou}. The attempts reconciling  the unpolarized and polarized measurements have mostly focused on improved treatments of radiative corrections (RC) \cite{Maximon,MASCARAD,TrRC,hadvar}, particularly on the theoretical estimation of the two-photon exchange contribution (see review \cite{Afanasev_review} and references therein). 

The interest to investigation of higher-order QED effects \cite{AGIM2015} was furthered by a so-called ''proton radius puzzle'' coming from the different outcomes of the
measurements in electron-proton systems \cite{CODATA,Sick} and in  the muonic hydrogen
\cite{Pohl}. The result of a recent experiment PRAD \cite{prad}  with  1.1 and 2.2~GeV electron beams is rather close to the muonium spectroscopy experiment and therefore disagrees with previous electron-proton scattering data.    
This unexpected issue required new efforts for theoretical and experimental investigations. Particularly, a MUSE experiment  is currently taking data at PSI \cite{MUSE} with incoming electron and muon beam momenta  115, 153, and 210 MeV, and PRAD-II using 3 different beam energies of 0.7, 1.4 and 2.1 GeV is in a planning stage at JLab \cite{PRAD2}. In theory, an unified treatment of both chiral and radiative corrections to the low-energy elastic lepton-proton scattering processes has been performed in Heavy Baryon Chiral Perturbations Theory \cite{Talukdar}.

In the loop integration with respect to the additional virtual particle momenta for the two-photon exchange sub-process, a virtuality of one photon tends to zero reaching a so-called soft limit that produces the infrared divergence.The latter should be canceled with the corresponding soft term from real photon emission. Since both the virtual and real soft photon terms are reduced to overall factors in front of the Born cross section independently of hadronic structure, in many cases real photon emission is estimated within the soft photon approximation only.

It should be noted that rather important uncertainties arise from emission of hard unobserved photons. The first systematic approach for the calculation of hard photon emission was presented by Mo and Tsai \cite{MoTsai1969}. Based on this approach Monte-Carlo generator ESEPP for hard photon simulation in elastic lepton- and antilepton-proton scattering \cite{ESEPP} was developed.  

However, one essential limitation in their calculations consists in the approximate approach for the treatment of the soft photon contribution. As a result, their final expressions depend on an artificial parameter $ \Delta$ that was introduced  to separate the photon momentum phase space into the ``soft" and ``hard" parts. In numerical calculations on one hand this parameter should be chosen as small as possible to reduce the region evaluated approximately, but on another hand it cannot be chosen too small because of possible numerical instabilities in calculating hard-photon emission. In 1977 Bardin and Shumeiko in their paper \cite{BSh} shown one of the solutions to the problem.

Recently, in our work \cite{chaslet} we demonstrated the influence of hard photon emission on the charge asymmetry in elastic lepton- and antilepton-proton scattering at fixed transferred momentum squared. This a charge-odd contribution comes from the interference of real photon emission from lepton and hadron legs. For cancellation of infrared divergence two-photon exchange is considered in soft photon approximation only. In the second paper \cite{lpcth}, a dramatic difference between RC with hard photon emission from the leptons in elastic $lp$-interaction for fixed $Q^2$ and scattering angle was demonstrated.

Here, we continue the studies from these papers by consideration of charge-odd hard photon emission presented in  \cite{chaslet} for two different kinematic constraints: either by fixing lepton's scattering angle or the transferred momentum squared. We also provide details of the calculations, explicit expressions as well as numerical comparison of charge asymmetries induced by real photon emission for two different observables, namely, $Q^2$ and a scattering angle.   Similar to the previous work \cite{MASCARAD,AGIM2015,chaslet,lpcth} the infrared divergence from real photon emission is separated by the Bardin-Shumeiko approach \cite{BSh}.  All calculations have been performed  beyond the ultrarelativistic limit, that allow to apply the obtained results for 
MUSE experiment \cite{MUSE} where a moderately relativistic muon beams are used.

Here, as in the previous work  \cite{chaslet} we will consider the two-photon exchange sub-process only within the soft photon approximation. 
It should be noted that there is some arbitrariness in the extraction of the infrared divergence: only asymptotic behavior at the low virtual/real photon energy is important. In practice, two conventions are commonly used for the infrared part of the two-photon exchange process. The first one is from Tsai \cite{Tsai1961} and has to be expressed through the three-points integrals while the second expression was presented in the work of Maximon and Tjon \cite{Maximon} and led to calculations of the four-points integrals. 
In this paper we will reproduce the detailed calculation of Tsai's expressions that was used for numerical analysis in  \cite{chaslet}. For this purpose we use  dimensional regularization without any discussion of the difference between Tsai and Maximon-Tjon approaches that can be found in the review \cite{Afanasev_review}.

Another relevant issue is evaluation of multi-soft photon contribution. Since at the lowest order soft photon radiation is factorized in front of Born contribution, it was suggested by Yennie, Frautschi and Suura \cite{YFS}  to use an exponential procedure for summing up soft photon emission in all orders. Shumeiko \cite{Shum} demonstrated that a similar multi-soft photon effect allows to avoid a divergence in the region of pion threshold for deep inelastic scattering. A more detailed analysis of the problem that justified the exponentiation procedure in QED collision processes may be found in Ref.~\cite{Passarino2001}. This contribution is important for elastic charged lepton-nucleon scattering, especially for tight kinematic cuts that select soft radiation. Soft-photon exponentiation is included in most of the experimental analysis procedures, while this article is focused on {\it hard-photon} emission that is often omitted in data analysis.
 
During our calculation we assume that there is no any excited states of the intermediated proton in two-photon exchange. As a result, the protonic
propagator is taken as an elementary fermionic one. The second assumption is that the on-shell proton vertex, 
\begin{eqnarray}
\Gamma_\mu (q)=\gamma_\mu F_1(-q^2)+\frac {i\sigma_{\mu \nu}q^\nu}{2M}F_2(-q^2),
\end{eqnarray}
works properly within off-shell region. Here $\sigma_{\mu \nu}=i[\gamma_\mu,\gamma_\nu]/2$,  $q$ is a four-momentum of the incoming photon.
The Dirac ($F_1$) and Pauli ($F_2$) form factors can be expressed through the electromagnetic ones:    
\begin{eqnarray}
F_1(-q^2)&=&\frac{G_E(-q^2)+\tau_p G_M(-q^2)}{1+\tau_p},
\nonumber\\
F_2(-q^2)&=&\frac{G_M(-q^2)-G_E(-q^2)}{1+\tau_p},
\end{eqnarray}
where $\tau_p=-q^2/4M^2$.

In order to avoid misunderstandings and make this article self-consistent, we have to rewrite some 
rather important equations from our previous works \cite{chaslet,lpcth}.

The rest of the paper is organized in a following way. In the next section Born contribution is considered. The explicit expressions for the soft photon extraction from two-photon exchange by Tsai method are presented in Section~\ref{sec3}. The contributions of the unobserved hard real photon emission both for fixed $Q^2$ and fixed angle are obtained in Section~\ref{sec4} with detailed discussion of infrared divergence extraction by the Bardin-Shumeiko approach. Numerical results using MUSE and JLab experimental conditions as an example can be found in Section~\ref{na}. A brief discussion and conclusions are presented in the last Section.  The details of three-point loop integral calculations responsible for the infrared divergence in two-photon exchange contribution are considered in~\ref{tpli}.  The explicit expressions for the physical quantities associated with
real photon emission  are presented in~\ref{thex}. 
The details of the soft real photon treatment can be found in \ref{ap2}.

\section{Born contribution}

Let us consider Born contribution  
to the unpolarized elastic $l^\pm p$ scattering:
\begin{eqnarray}
l^\pm(k_1)+p(p_1)\longrightarrow l^\pm(k_2)+p(p_2), 
\label{nrad}
\end{eqnarray}
where $k_1$, $p_1$ ($k_2$, $p_2$) are the initial (final) lepton and proton four-momenta respectively ($k_1^2=k_2^2=m^2$, $p_1^2=p_2^2=M^2$). Despite we consider this process in the target rest frame reference system (${\bf p_1}=0$), it will be useful to introduce some invariants:
\begin{eqnarray}
&\displaystyle
S=2p_1k_1,\; Q^2=-(k_1-k_2)^2=-q^2,\; X=S-Q^2,
\nonumber\\[2mm]
&\displaystyle
\lambda_S=S^2-4m^2M^2,
\lambda_X=X^2-4m^2M^2.
\end{eqnarray}

\begin{figure}[t]
\hspace*{2mm}
\includegraphics[width=3.5cm,height=4cm]{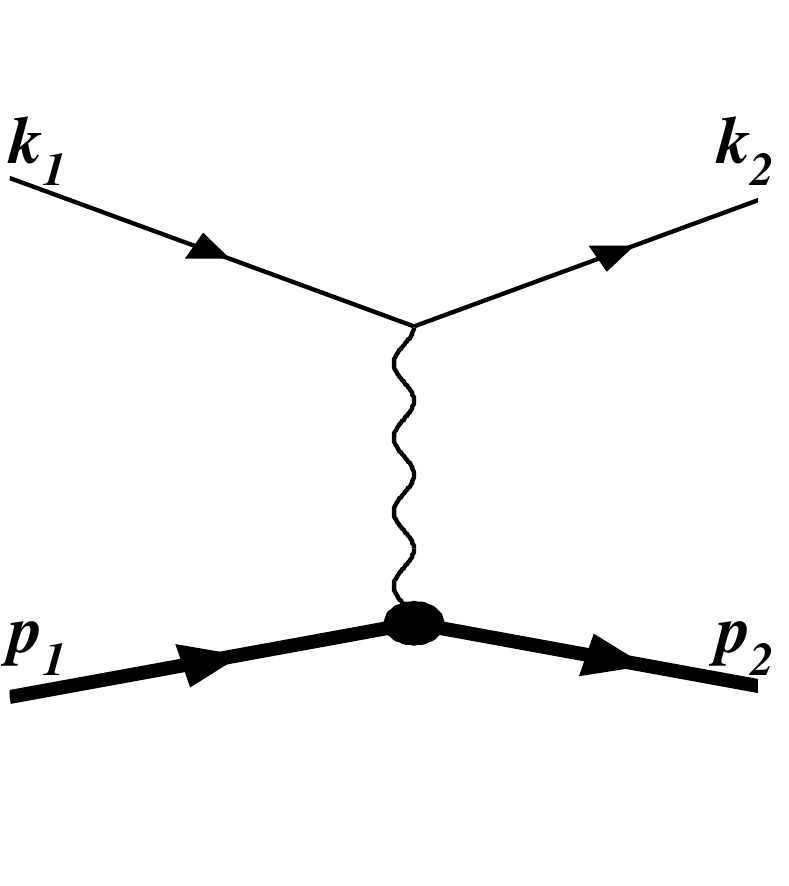}
\hspace*{5mm}
\includegraphics[width=3.5cm,height=4cm]{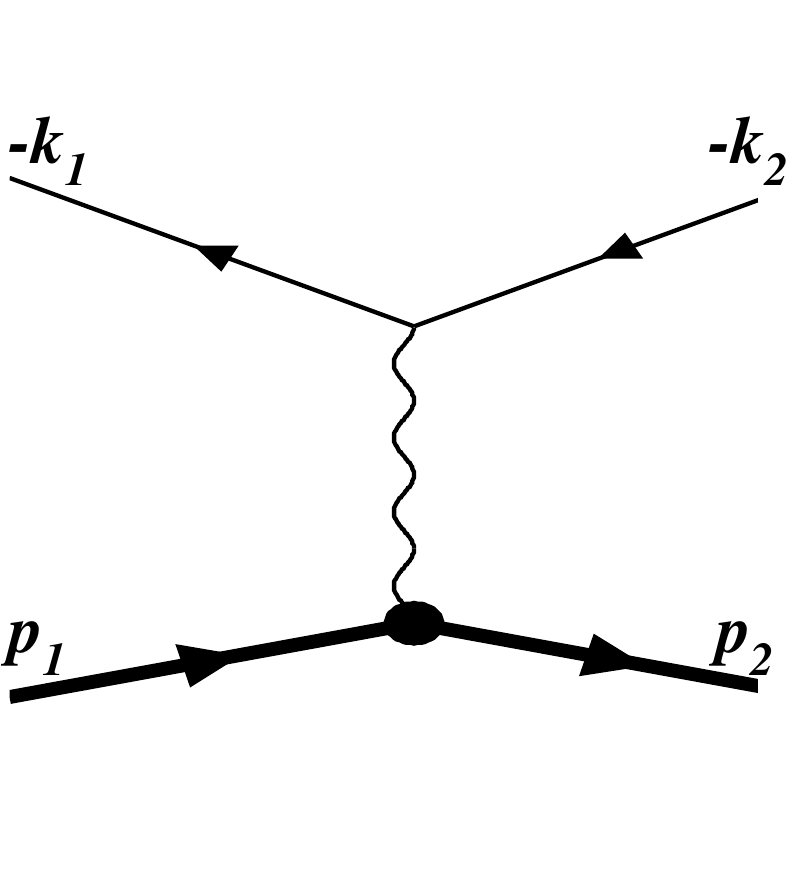}
\put(-174,10){\mbox{a)}}
\put(-52,10){\mbox{b)}}
\caption{Feynman graphs for the lowest-order
contribution to elastic $l^-p$ (a) and $l^+p$ (b) scattering.
}
\label{fig1}
\end{figure}
Similar to our previous work \cite{lpcth} in the present paper we will interest in two types of the cross section: $d\sigma/dQ^2$ and $d\sigma/d\cos\theta$. 

The considered contribution is
presented by Feynman graphs in Fig.~\ref{fig1} and can be described by the following matrix elements:  
\begin{eqnarray}
{\cal M}^-_b&=&\frac{ie^2}{Q^2}
{\bar u}(k_2)\gamma^\mu u(k_1)
{\bar U}(p_2)\Gamma_\mu (q)U(p_1),
\nonumber\\
{\cal M}^+_b&=&\frac{ie^2}{Q^2}
{\bar u}(-k_1)\gamma^\mu u(-k_2)
{\bar U}(p_2)\Gamma_\mu (q)U(p_1),
\end{eqnarray}
where $e=\sqrt{4\pi\alpha}$.
Since the squares of these two matrix elements are identical it is not possible to distinguish
the lepton-proton from antilepton-proton scattering processes on the one-photon exchange
(Born) level. Their contributions to the cross section are even regarding replacement $l^-\leftrightarrow l^+$ 
and can be written similar to \cite{lpcth}
\begin{eqnarray}
d\sigma_B&=&\frac 1{2\sqrt{\lambda_S}}|{\cal M}^\mp_b|^2 d\Gamma_2, 
\end{eqnarray}
where  the phase space looks like:  
\begin{eqnarray}
d\Gamma_2&=&(2\pi)^4\delta ^4(p_1+k_1-p_2-k_2)
\frac{d^3k_2}{(2\pi)^32k_{20}}
\frac{d^3p_2}{(2\pi)^32p_{20}}
\nonumber\\
&=&\frac{dQ^2}{8\pi\sqrt{\lambda_S}}
=\frac{\sqrt{\lambda_X}d\cos \theta}{8\pi(S+2M^2-\cos \theta X\sqrt{\lambda_S/\lambda_X})}.
\end{eqnarray}

After simplest calculations for our future purposes Born contribution to the cross section can be presented 
in different way with respect to \cite{lpcth}:
\begin{eqnarray}
\frac{d\sigma_B}{dQ^2}&=&\frac{2\pi \alpha^2}{\lambda_S Q^4}\sum_{i,j=1}^2\theta _{ij}^BF_i(Q^2)F_j(Q^2),  
\nonumber\\
\frac{d\sigma_B}{d\cos\theta}&=&j_\theta\frac{d\sigma_B}{dQ^2},
\end{eqnarray}
where 
\begin{eqnarray}
j_\theta=\frac{\sqrt{\lambda_S}\lambda_X^{3/2}}{2M^2 (S X-2m^2(Q^2+2M^2))},
\end{eqnarray}
and the quantities $\theta _{ij}^B$ have a form:
\begin{eqnarray}
\theta _{11}^B&=&S^2+X^2-2Q^2(M^2+m^2),
\nonumber\\
\theta _{12}^B&=&\theta _{21}^B=Q^2(Q^2-2m^2),  
\nonumber\\
\theta _{22}^B&=&\frac {Q^2}{2M^2}(SX+M^2(Q^2-4m^2)).  
\end{eqnarray}

\section{Soft photon extraction from two-photon exchange}
\label{sec3}
\begin{figure}[t]
\hspace*{2mm}
\includegraphics[width=3.5cm,height=4cm]{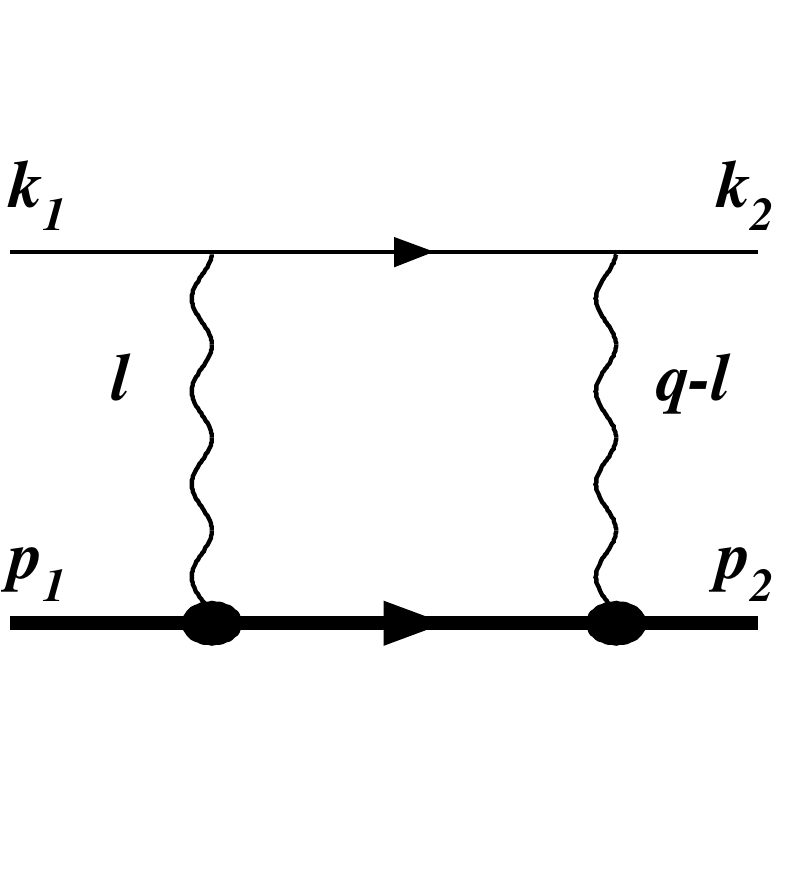}
\hspace*{5mm}
\includegraphics[width=3.5cm,height=4cm]{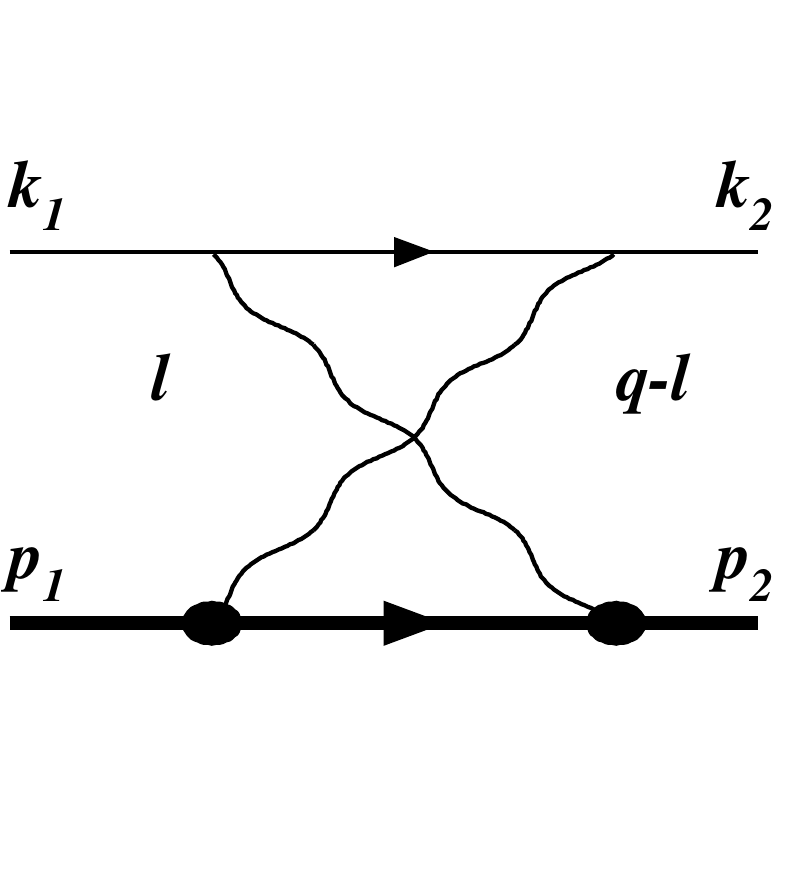}
\put(-170,10){\mbox{(a)}}
\put(-50,10){\mbox{(b)}}
\vspace*{-2mm}
\caption{Feynman graphs for the direct (a) and
cross (b) two-photon exchange   within $l^-p$-scattering. The similar graphs for $l^+p$ scattering processes have
an opposite direction  for the leptonic arrows and a negative sign for its momenta.
}
\label{fig2}
\end{figure}

The matrix elements with the
two-photon exchange contribution in elastic $l^\mp p$ scattering can be
separated into the direct ${\cal M}^\mp_{2\gamma d}$ and cross ${\cal M}^\mp_{2\gamma x}$ terms, as it is depicted in Fig.~\ref{fig2}(a) and Fig.~\ref{fig2}(b) respectively. Each of these contributions
can be presented through the loop integration as it was shown in our previous work \cite{chaslet}:
\begin{strip}
\rule[-1ex]{\columnwidth}{1pt}\rule[-1ex]{1pt}{1.5ex}
\begin{eqnarray}
{\cal M}^-_{2\gamma d}&=&\frac{e^4}{(2\pi)^4}\int\frac{d^4l}{l^2(l-q)^2}
{\bar u}(k_2)\gamma^\nu\frac{\slashed k_1-\slashed l+m}{l^2-2k_1l}\gamma^\mu u(k_1)
{\bar U}(p_2)\Gamma_\nu (q-l)\frac{\slashed p_1+\slashed l+M}{l^2+2p_1l}\Gamma_\mu(l)U(p_1),
\nonumber\\
{\cal M}^+_{2\gamma d}&=&\frac{e^4}{(2\pi)^4}\int\frac{d^4l}{l^2(l-q)^2}
{\bar u}(-k_1)\gamma^\mu\frac{\slashed l-\slashed k_1+m}{l^2-2k_1l}\gamma^\nu u(-k_2)
{\bar U}(p_2)\Gamma_\nu (q-l)\frac{\slashed p_1+\slashed l+M}{l^2+2p_1l}\Gamma_\mu(l)U(p_1),
\nonumber\\
{\cal M}^-_{2\gamma x}&=&\frac{e^4}{(2\pi)^4}\int\frac{d^4l}{l^2(l-q)^2}
{\bar u}(k_2)\gamma^\nu\frac{\slashed k_1-\slashed l+m}{l^2-2k_1l}\gamma^\mu u(k_1)
{\bar U}(p_2)\Gamma_\mu (l)\frac{\slashed p_2-\slashed l+M}{l^2-2p_2l}\Gamma_\nu(q-l)U(p_1),
\nonumber\\
{\cal M}^+_{2\gamma x}&=&\frac{e^4}{(2\pi)^4}\int\frac{d^4l}{l^2(l-q)^2}
{\bar u}(-k_1)\gamma^\mu\frac{\slashed l-\slashed k_1+m}{l^2-2k_1l}\gamma^\nu u(-k_2)
{\bar U}(p_2)\Gamma_\mu (l)\frac{\slashed p_2-\slashed l+M}{l^2-2p_2l}\Gamma_\nu(q-l)U(p_1).
\label{m2g}
\end{eqnarray}
\hfill\rule[1ex]{1pt}{1.5ex}\rule[2.3ex]{\columnwidth}{1pt}
\end{strip}

The interference of these matrix elements with Born ones
\begin{eqnarray}
d\sigma_{2\gamma }^\mp&=&\frac 1{2\sqrt{\lambda_S}}[
{\cal M}^\mp_b 
({\cal M}^\mp_{2\gamma d}+{\cal M}^\mp_{2\gamma x})^\dagger 
\nonumber\\&&
+({\cal M}^\mp_{2\gamma d}+{\cal M}^\mp_{2\gamma x})
{\cal M}^{\mp\; \dagger}_b ] 
d\Gamma_2 
\label{s2g}    
\end{eqnarray}
give odd regarding replacement $l^-\leftrightarrow l^+$ contribution 
to the elastic $l^\mp p$ cross section 
\begin{eqnarray}
d\sigma_{2\gamma }^+=-d\sigma_{2\gamma }^-.
\end{eqnarray}

However, as it was mentioned in Introduction, from the presented above contribution we are interested here only in the soft photon parts coming from the situation when one of two-photons has a low virtuality  
at $l\to 0$ or $l\to q$. These terms are rather important since they contain the infrared divergence that has to be canceled with corresponding divergence from the real photon emission.

Therefore in each of the matrix elements from (\ref{m2g}) taking into account $F_1(0)=1$
we can extract two contributions following Tsai suggestion \cite{Tsai1961}:
\begin{eqnarray}
{\cal M}^\mp_{2\gamma \{d,x\} \;IR}&=&
{\cal M}^\mp_{2\gamma \{d,x\} }\Big|_{l\to 0}+
{\cal M}^\mp_{2\gamma \{d,x\} }\Big|_{l\to q}.
\end{eqnarray}
Here 
\begin{eqnarray}
{\cal M}^\mp_{2\gamma d}\Big|_{l\to 0}&=&\mp\frac {e^2}{8\pi^2}{\cal M}^\mp_b K_{IR}(k_1,-p_1),
\nonumber\\
{\cal M}^\mp_{2\gamma d}\Big|_{l\to q}&=&\mp\frac {e^2}{8\pi^2}{\cal M}^\mp_b K_{IR}(k_2,-p_2),
\nonumber\\
{\cal M}^\mp_{2\gamma x}\Big|_{l\to 0}&=&\pm\frac {e^2}{8\pi^2}{\cal M}^\mp_b K_{IR}(k_1,p_2),
\nonumber\\
{\cal M}^\mp_{2\gamma x}\Big|_{l\to q}&=&\pm\frac {e^2}{8\pi^2}{\cal M}^\mp_b K_{IR}(k_2,p_1).
\end{eqnarray}
The detailed calculation of the infrared three-point integrals
\begin{eqnarray}
K_{IR}(a,b)=-\frac {2ab}{i\pi^2}\int\frac{(2\pi \mu)^{4-n}d^nl}{l^2(l^2-2al)(l^2-2bl)},
\label{kab}
\end{eqnarray}
where $\mu$ is a free parameter of a mass dimension presented in~\ref{tpli} shows that 
$K_{IR}(k_1,-p_1)=K_{IR}(k_2,-p_2)$ are complex quantities while $K_{IR}(k_1,p_2)=K_{IR}(k_2,p_1)$ are real ones.

After replacing ${\cal M}^\mp_{2\gamma \{d,x\}}\to {\cal M}^\mp_{2\gamma \{d,x\} \;IR}$
in Eq.~(\ref{s2g}) one can find that the real part of soft photon contribution extracted from 
two-photon exchange contains the infrared divergence $P_{IR}$ defined by Eq.~(\ref{pir}): 
\begin{eqnarray}
\frac{d\sigma_{2\gamma}^{IR\;\mp}}{dQ^2}
&=&\mp\frac{\alpha }{\pi}
\biggl({\rm Re}[K_{IR}(k_1,-p_1)+K_{IR}(k_2,-p_2)]
\nn&&
-K_{IR}(k_1,p_2)-K_{IR}(k_2,p_1)\biggr)\frac{d\sigma_B}{dQ^2}
\nonumber\\
&=&\mp\frac{\alpha }{\pi}\delta^{IR} _{2\gamma}(S,X)
\frac{d\sigma_B}{dQ^2}.
\label{irb0}
\end{eqnarray}

As it was shown in Section 4 of our previous work \cite{lpcth} the calculation of the additional virtual particle contributions to the leptonic current come to the ultraviolet divergence. The latter can be removed by virtue of shifting a constant factor when the exchange photon virtuality is fixed {\it i.e. } $Q^2=Q^2_0$. In many cases, such as \cite{lpcth}, for which $Q^2_0=0$, this subtraction procedure is called an on-shell (mass-shell) renormalization scheme. 

Despite the fact that Eq.~(\ref{irb0}) does not contain the ultraviolet divergence a physical requirement of on-shell renormalization scheme comes from vanishing asymmetry at $Q^2\to 0$ can be provided by the difference $\delta_{2\gamma}^{IT}(S,X)$ and its value at $X=S$ (or $Q^2=0$).
As a results the required contribution has a form:
\begin{eqnarray}
\frac{d\hat \sigma_{2\gamma}^{IR\;\mp}}{dQ^2}
&=&\mp\frac{\alpha }{\pi}\hat\delta^{IR} _{2\gamma}
\frac{d\sigma_B}{dQ^2},
\label{irb}
\end{eqnarray}
where
\begin{eqnarray}
\hat\delta^{IR}_{2\gamma}&=&\delta^{IR}_{2\gamma}(S,X)-\delta^{IR}_{2\gamma}(S,S)
\nonumber\\
&=&2(SL_S-XL_X)
\left(P_{IR}+\log\frac m\mu
\right)+\delta_{2\gamma},
\end{eqnarray}
and
\begin{eqnarray}
L_S&=&\frac 1{\sqrt{\lambda_S}}\log\frac {S+\sqrt{\lambda_S}} {S-\sqrt{\lambda_S}}, 
\nonumber\\
L_{X}&=&\frac 1{\sqrt{\lambda_{X}}}\log\frac {X+\sqrt{\lambda_{X}}} {X-\sqrt{\lambda_{X}}}.
\label{lslx} 
\end{eqnarray} 
The infrared free part reads:
\begin{eqnarray}
&\displaystyle
\delta_{2\gamma}=
\frac S{\sls}\Biggl[
2{\rm Li}_2\frac{\sls-S+2M^2}{2\sls}
\nonumber\\
&\displaystyle
+
\log\frac{2\lambda_S(S+\sls)}{m^2(S-2M^2+\sls)^2}
\log\frac{S+\sls}{2M^2}
\nonumber\\
&\displaystyle
+\log \frac M m \log\frac {S+\sls}{S-\sls}
-2{\rm Li}_2\frac{S-2m^2+\sls}{2\sls}
\Biggr]
\nonumber\\
&\displaystyle
+\frac X{\slx}\Biggl[
2{\rm Li}_2\frac{X+\slx-2m^2}{2\slx}
-\frac 12\log^2\frac {X+\slx}{X-\slx}
\nonumber\\&\displaystyle
+
\log\frac{(X+\slx)(X-2M^2+\slx)^2}{8M^2\lambda_X}
\log\frac{X+\slx}{2M^2}
\nonumber\\&\displaystyle
\nonumber\\
&\displaystyle
-2{\rm Li}_2\frac{2M^2+\slx-X}{2\slx}
\Biggr]
.
\end{eqnarray}

Notice, that in ultrarelativistic approximation for $m\to~0$
\begin{eqnarray}
\hat\delta^{IR}_{2\gamma}&=&
\Biggl[4P_{IR}+\log \frac{SX}{\mu^4}
\Biggr]\log \frac S X
-2{\rm Li}_2 \Biggl[1-\frac{M^2}S\Biggr]
\nonumber\\&&
+2{\rm Li}_2 \Biggl[1-\frac{M^2}X\Biggr].
\end{eqnarray}
Taking into account that $P_{IR}=\log (\mu/\lambda )$, $S=2ME$ and $X=2ME^\prime$ we immediately find the agreement of the obtained result with Eq.~(2.20) of the review \cite{Afanasev_review}.

\section{Real photon contribution}
\label{sec4}
\begin{figure}[t]
\hspace*{2mm}
\includegraphics[width=3.5cm,height=4cm]{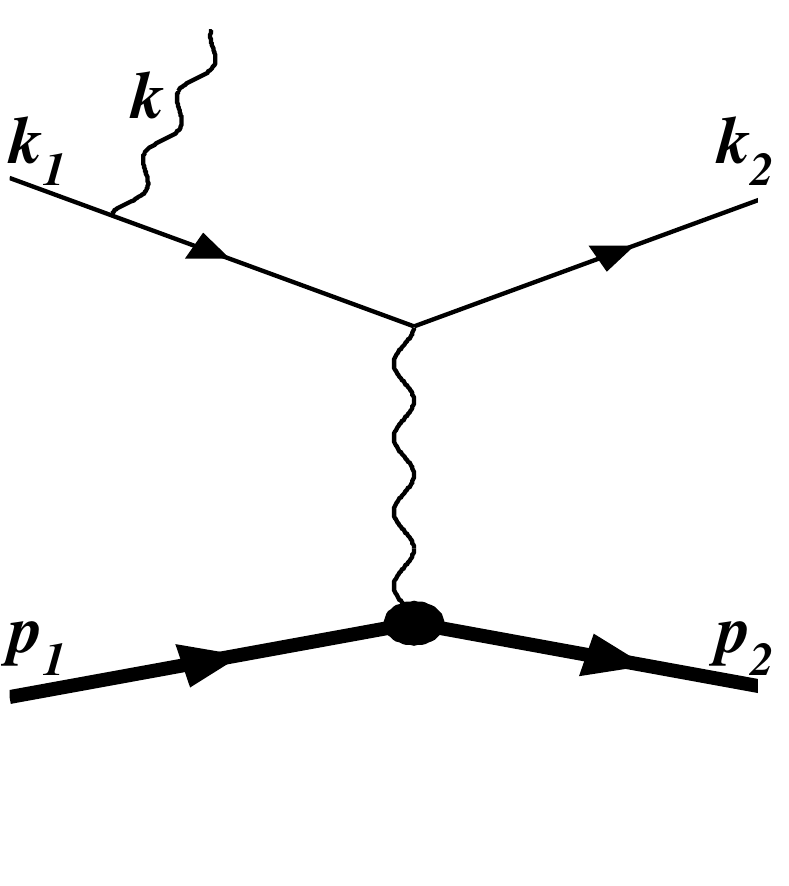}
\hspace*{5mm}
\includegraphics[width=3.5cm,height=4cm]{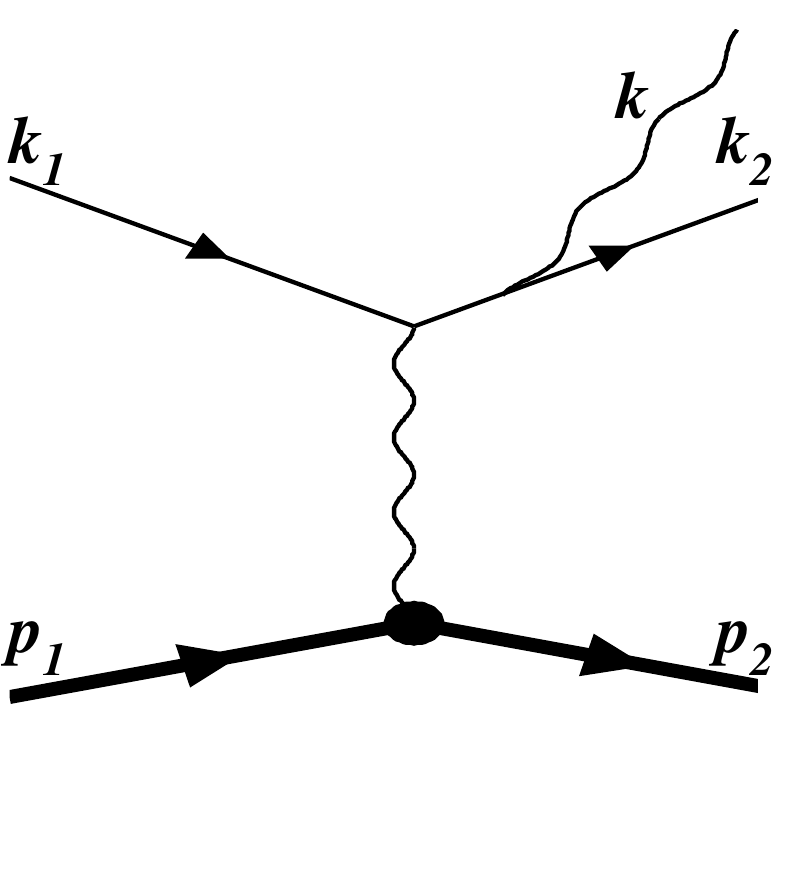}
\put(-150,10){\mbox{a)}}
\put(-48,10){\mbox{b)}}
\\[-5mm]
\hspace*{2mm}
\includegraphics[width=3.5cm,height=4cm]{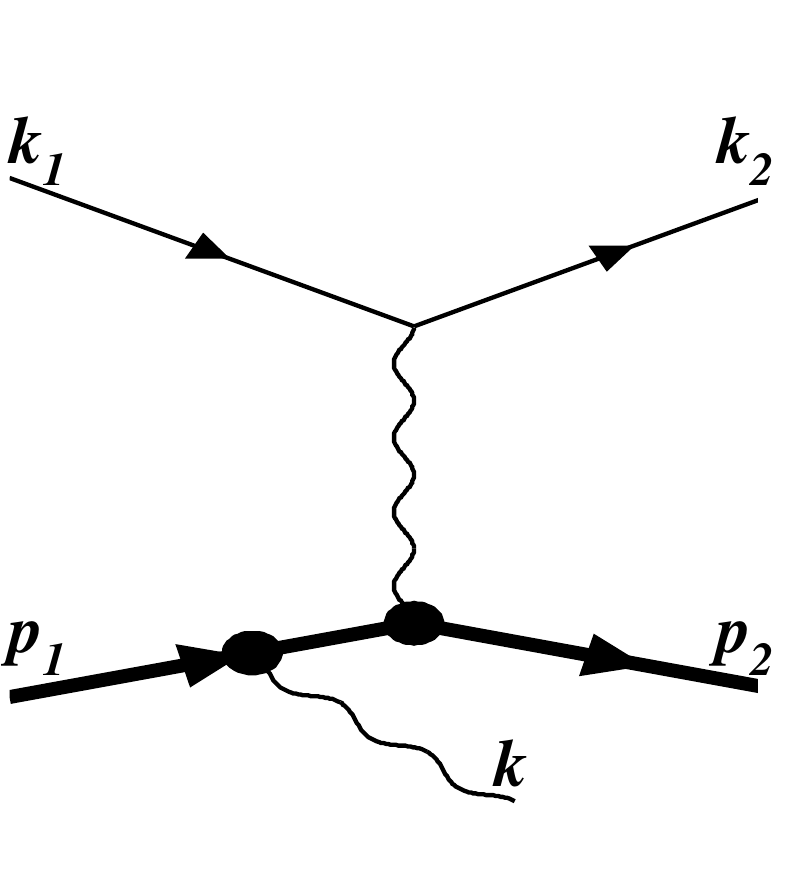}
\hspace*{5mm}
\includegraphics[width=3.5cm,height=4cm]{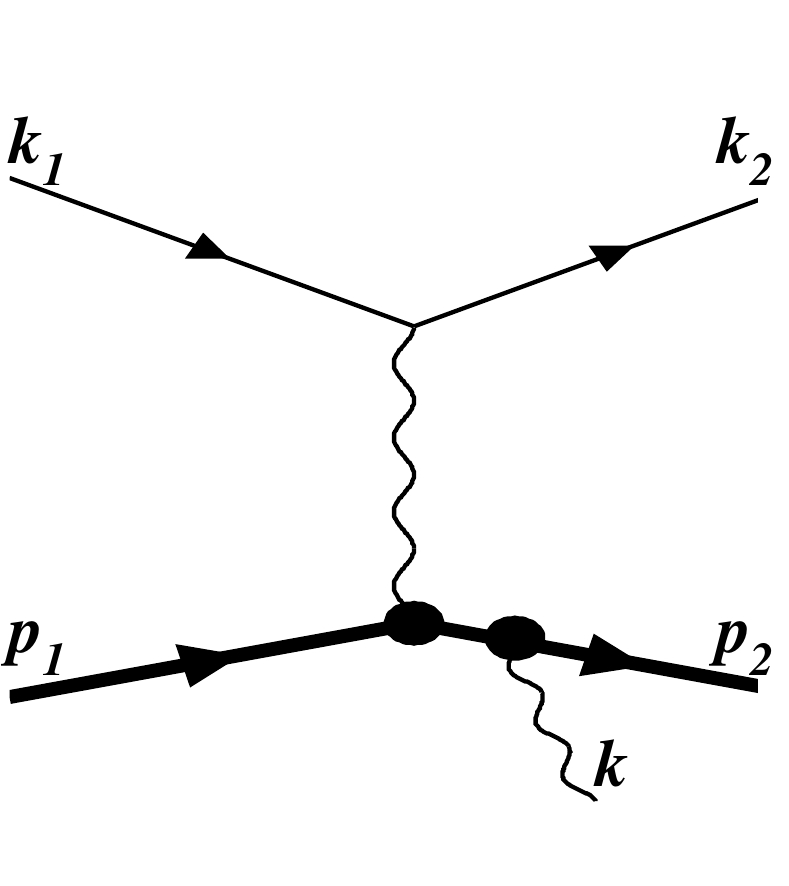}
\put(-150,0){\mbox{c)}}
\put(-48,0){\mbox{d)}}
\caption{Feynman graphs for real photon emission from the lepton (a,b) and proton (c,d) legs within $l^-p$-scattering. The similar graphs for $l^+p$ scattering processes have
an opposite direction  for the leptonic arrows and a negative sign for its momenta.
}
\label{fig3}
\end{figure}

In this section the odd regarding replace $l^-\leftrightarrow l^+$ contribution
from real photon emission
\begin{eqnarray}
l^\pm(k_1)+p(p_1)\longrightarrow l^\pm(k_2)+p(p_2)+k(k) 
\label{rad}
\end{eqnarray}
to the unpolarized elastic $l^\pm p$ scattering is considered. It consists of the interference between the matrix elements with the real photon emission from the lepton and proton legs as it is presented in Fig.~\ref{fig3}. 

For the description of real photon emission three additional variables has to be introduced. We choose the standard set \cite{MASCARAD,lpcth} of them: inelasticity $v=(p_1+k_1-k_2)^2-M^2$, 
$\tau=kq/kp_1 $ and  $\phi_k$ is an angle between
(${\bf k}_1$,${\bf k}_2$) and (${\bf k}$,${\bf q}$) planes in the rest frame (${\bf p}_1=0$).

As it was shown in \cite{lpcth} the upper kinematical limit over the inelasticity
depends on fixed variable. Namely, for fixed $Q^2$ the maximum inelasticity value is
\begin{eqnarray}
v_q=\frac {\sqrt{\lambda_S}\sqrt{Q^2(Q^2+4m^2)}-2m^2Q^2-Q^2S}{2m^2},
\label{vmq2}
\end{eqnarray}
while for the fixed scattering angle
\begin{eqnarray}
v_{\theta}=S+2m^2-
\frac m M \sqrt{(S+2M^2)^2-\lambda_S\cos^2\theta}.
\end{eqnarray}
On the other hand, the contribution of hard real photon emission can be controlled
by applying a cut $v_{cut}$ on the inelasticity in the single-arm
measurement of the scattered lepton only. Therefore, keeping in mind the inelasticity maximum values, for an upper limit of  this quantity  both for fixed $Q^2$ and scattering angle we will use $v_{cut}$ as an experimentally observable variable.

The matrix elements corresponding to real photon emission in $l^\mp p$ scattering have a form:
\begin{eqnarray}
{\cal M}^{-}_{lR}
&=&-\frac{ie^3}t
{\bar u}(k_2) \varepsilon_\alpha \Gamma_{lR}^{\mu \alpha} u(k_1)
{\bar U}(p_2)\Gamma_\mu(q-k)U(p_1),
\nonumber\\
{\cal M}^{+}_{lR}
&=&\frac{ie^3}t
{\bar u}(-k_1) \varepsilon_\alpha {\bar \Gamma}_{lR}^{\mu \alpha} u(-k_2)
{\bar U}(p_2)\Gamma_\mu(q-k)U(p_1),
\nonumber\\
{\cal M}^{-}_{hR}
&=&\frac{ie^3}{Q^2}{\bar u}(k_2)\gamma_\mu u(k_1){\bar U}(p_2)\varepsilon_\alpha \Gamma_{hR}^{\mu \alpha}U(p_1),
\nonumber\\
{\cal M}^{+}_{hR}
&=&\frac{ie^3}{Q^2}{\bar u}(-k_1)\gamma_\mu u(-k_2){\bar U}(p_2)\varepsilon_\alpha \Gamma_{hR}^{\mu \alpha}U(p_1),
\label{mhlr}
\end{eqnarray}
where $t=-(q-k)^2=Q^2+\tau R$, $R=2p_1k=v/(1+\tau )$, $\varepsilon_\alpha$ is the photon polarized vector, and
\begin{eqnarray}
\Gamma_{lR}^{\mu \alpha}&=&\Biggl(
\frac{k_{1\alpha }}{kk_1}- 
\frac{k_{2\alpha }}{kk_2}\Biggr)\gamma^\mu
-\frac{\gamma^\mu \hat k \gamma^\alpha}{2k_1k} 
-\frac{ \gamma^\alpha\hat k \gamma^\mu}{2k_2k},
\nonumber\\
{\bar \Gamma}_{lR}^{\mu \alpha}&=&\Biggl(
\frac{k_{1\alpha }}{kk_1}- 
\frac{k_{2\alpha }}{kk_2}\Biggr)\gamma^\mu
-\frac{ \gamma^\alpha\hat k \gamma^\mu}{2k_1k}
-\frac{\gamma^\mu \hat k \gamma^\alpha}{2k_2k}, 
\nonumber\\
\Gamma_{hR}^{\mu \alpha}&=&
\Gamma^\mu(q)\frac{\slashed p_1-\slashed k+M}{2p_1k}\Gamma^\alpha (-k)
\nonumber\\
&&  
-\Gamma^\alpha (-k) 
\frac{\slashed p_2+\slashed k+M}{2p_2k}\Gamma^\mu(q).
\end{eqnarray}

The part of the cross section with 
the interference between real photon emission from hadron and lepton lines reads:
\begin{eqnarray}
d\sigma_R^\mp=\frac 1{2\sqrt{\lambda_S}}({\cal M}^\mp_{lR}{\cal M}^{\mp\;\dagger}_{hR}+{\cal M}^\mp_{hR}{\cal M}^{\mp\;\dagger}_{lR})
d\Gamma_3,
\label{sri}
\end{eqnarray}
where the phase space has a form:
\begin{eqnarray}
d\Gamma_3&=&
\frac{d^3k}{(2\pi)^32k_{0}}
\frac{d^3k_2}{(2\pi)^32k_{20}}
\frac{d^3p_2}{(2\pi)^32p_{20}}
\nonumber\\&&
\times
(2\pi)^4\delta ^4(p_1+k_1-p_2-k_2-k).
\end{eqnarray}
One can verify directly that the interference terms for $l^-p$ and $l^+p$ 
have an opposite sign
\begin{eqnarray}
d\sigma_R^+=-d\sigma_R^-.
\end{eqnarray}

\subsection{Contribution to $d\sigma/dQ^2$}
The phase space for fixed $Q^2$ can be presented as:
\begin{eqnarray}
d\Gamma_3&=&
\frac{dQ^2vdv d\tau d\phi_k}{2^8\pi^4(1+\tau)^2\sqrt{\lambda_S \lambda_q}}.
\end{eqnarray}
Here $\lambda_q=(Q^2+v)^2+4M^2Q^2$, and the radiative cross section can be decomposed in terms of variable $R$ introduced above:  
\begin{eqnarray}
d\sigma_R^\mp&=&\mp\frac{\alpha ^3dQ^2dvd\tau }{2Q^2t(1+\tau)\lambda_S}
\sum_{i,j,k=1}^2\sum_{l=l^d_{ijk}}^{l^u_{ijk}} 
\theta_{ijk}^l(Q^2,\tau)R^{l-2}
\nonumber\\&&\times
F_i(0)F_j(Q^2)F_k(t).
\label{sri1}
\end{eqnarray}
The lowest and upper limits in the sum over $l$ read:
\begin{eqnarray}
&&l^d_{1jk}=1, 
\nonumber\\
&&l^d_{212}=l^d_{221}=2,  
\nonumber\\
&&l^d_{211}=l^d_{222}=3,
\nonumber\\
&&l^u_{111}=l^u_{112}=l^u_{121}=l^u_{211}=4,
\nonumber\\
&&l^u_{122}=l^u_{221}=l^u_{212}=l^u_{222}=5.
\end{eqnarray}

Explicit expressions for $\displaystyle \theta_{ijk}^l(Q^2,\tau)$ integrated over $\phi_k$ can
be found in~\ref{thex}.


In order to estimate this contribution to the elastic process, it is  necessary to integrate
$d\sigma_R^\mp$ over three  variables of the unobservable photon: $v$, $t$ and $\phi_k$. 
However, since the expressions~(\ref{sri}) contain the infrared divergence  
at $i=l=1$ and $v=0$ (or $R=0$), straightforward integration is not possible.

This infrared contribution has to be extracted but in a rather arbitrary way keeping the same asymptotic behavior at $v \to 0$. In order to treat this divergence analytically
following the Bardin-Shumeiko approach \cite{BSh}, 
in the $i=l=1$ term the variable $t$ should be changed to $Q^2$ both in the photon propagator and
the argument of the form factors. As a result this infrared term factorizes in front of the Born cross section:   
\begin{eqnarray}
d\sigma_R^{IR\; \mp}&=&\mp\frac{\alpha }{\pi^2}d\sigma_B \frac {vdvd\tau d\phi_k }{2(1+\tau)^2\sqrt{\lambda_q}}{\cal F}_{IR},
\label{irr1}
\end{eqnarray}
where
\begin{eqnarray}
{\cal F}_{IR}&=&-\frac12
\Biggl(
\frac{k_{1\alpha}}{k_1k}-
\frac{k_{2\alpha}}{k_2k}
\Biggr)
\Biggl(
\frac{p_{1\alpha}}{p_1k}-
\frac{p_{2\alpha}}{p_2k}
\Biggr).
\label{fir}
\end{eqnarray}

 This replacement allows us to perform the treatment of the infrared divergence analytically
since the arguments of the form factors in (\ref{irr1}) do not depend on photonic
variables.
For this purpose it is necessary to separate the factorized infrared term into the soft $\delta_S$ and hard $\delta_H$ parts
\begin{eqnarray}
\frac {d\sigma_R^{IR\; \mp}}{dQ^2}=\mp \frac \alpha \pi (\delta_S+\delta_H)\frac {d\sigma_{B}}{dQ^2}
\end{eqnarray}
by introducing of the infinitesimal inelasticity $\bar v$
\begin{eqnarray}
\delta_S&=&\frac 1\pi\int\limits^{\bar v}_0dv\int \frac{d^3k}{k_0}\delta((p_1+q-k)^2-M^2){\cal F}_{IR},
\nonumber\\
\delta_H&=&\frac 1\pi\int\limits_{\bar v}^{v_{cut}}dv\int \frac{d^3k}{k_0}\delta((p_1+q-k)^2-M^2){\cal F}_{IR}.\;\;
\label{dsh}
\end{eqnarray}
Such separation allows us to calculate $\delta_H$ for
$n=4$
and to simplify the integration for $\delta_S$ in the dimensional regularization by choosing the individual reference systems for each invariant variable
to make them independent of the azimuthal angle $\phi_k$.

The analytical integration of $\delta_S$ using the dimensional regularization
-- as presented in \ref{ap2} -- gives
\begin{eqnarray}
&&\delta_S=2(XL_X-SL_S)\biggl(P_{IR}+\log \frac {\bar v}{\mu M}\biggr)
+\frac S4 \sqrt{\lambda_S}L_S^2
\nonumber\\[1mm]&&\qquad
-\frac X4 \sqrt{\lambda_X}L_X^2
+\frac S{\sqrt{\lambda_S}}{\rm Li}_2\biggl(\frac {2\sqrt{\lambda_S}}{S+\sqrt{\lambda_S}}\biggr)
\nonumber\\[1mm]&&
\qquad
-\frac X{\sqrt{\lambda_X}}{\rm Li}_2\biggl(\frac {2\sqrt{\lambda_X}}{X+\sqrt{\lambda_X}}\biggr)
+S_\phi(k_2,p_1,p_2)
\nonumber\\[2mm]&&
\qquad
-S_\phi(k_1,p_1,p_2),
\end{eqnarray}
where the infrared divergent term $P_{IR}$ is defined by Eq.~(\ref{pir})
and the arbitrary parameter $\mu $ has a dimension of mass.
A rather compact analytical expression for $S_\phi $ can be found in Appendix B of the work \cite{lpcth}.

For the calculation of $\delta_H$ we 
integrate directly with respect to the photonic variables $v$, $t$ and $\phi_k$.
The first integration over the azimuthal angle gives:
\begin{eqnarray}
\delta_H
&=&\int\limits_{\bar v}^{v_{cut}}dv\int\limits_{\tau^q_{min}}^{\tau^q_{max}}\frac{d\tau}{1+\tau}\frac{ F_{IR}}{R},
\label{dh}
\end{eqnarray}
where the limits of variable $\tau $ are defined as in \cite{lpcth}
\begin{eqnarray}
\tau^q_{max/min}=\frac{Q^2+v\pm\sqrt{\lambda_q}}{2M^2},
\label{ta}
\end{eqnarray}
and the expression for $F_{IR}$ is presented in Eq.~(\ref{fff}).

Since $R=v/(1+\tau)$
the second integration over $\tau$ gives:
\begin{eqnarray}
\delta_H
&=&\int\limits_{\bar v}^{v_{cut}}\frac{dv}v J_H(S,X,v),
\label{dh2}
\end{eqnarray}
where
\begin{eqnarray}
J_H(S,X,v)&=&
\frac X{\gamma_3}\log\frac{X-v+\gamma_3}{X-v-\gamma_3}
+
\frac X{\gamma_2}\log\frac{X+\gamma_2}{X-\gamma_2}
\nonumber\\&&
-SL_S
-\frac S{\gamma_1}\log\frac{S-v+\gamma_1}{S-v-\gamma_1},
\label{jh}
\end{eqnarray}
and
\begin{eqnarray}
\gamma_1&=&\sqrt{(S-v)^2-4m^2(M^2+v)},
\nonumber\\
\gamma_2&=&\sqrt{X^2-4m^2(M^2+v)},
\nonumber\\
\gamma_3&=&\sqrt{(X-v)^2-4m^2M^2}.
\end{eqnarray}

Taking into account that
\begin{eqnarray}
J_H(S,X,0)&=&2(XL_X-SL_S),
\label{dhb}
\end{eqnarray}
after an identical transformation, Eq.~(\ref{dh2}) can be split into two parts
$\delta_H=\delta_H^1+\bar \delta_H$ with
\begin{eqnarray}
\delta_H^1&=&
\int\limits_{\bar v}^{v_{cut}}\frac{dv}v (J_H(S,X,v)-J_H(S,X,0)),
\nonumber\\
\bar \delta_H&=&2(XL_X-SL_S)\log\frac {v_{cut}}{\bar v}, 
\end{eqnarray}
where the integrand in $\delta_H^1$ is finite for $v\to 0$.
Since, as mentioned above, the arbitrariness in a choice of a subtracted infrared expression
is only constrained by its asymptotic behavior at $v\to 0$,
we can drop $\delta_H^1$ term. 

As a result, the contribution with the real photon emission can be split into finite and infrared divergent parts:
\begin{eqnarray}
\frac {d\sigma_R^{\mp}}{dQ^2}&=&
\frac {d\sigma_R^{\mp}}{dQ^2}
-\frac {d\bar\sigma_R^{IR\; \mp}}{dQ^2}
+\frac {d\bar\sigma_R^{IR\; \mp}}{dQ^2}
\nonumber\\
&=&
\frac {d\sigma_F^{\mp}}{dQ^2}
+\frac {d\bar\sigma_R^{IR\; \mp}}{dQ^2}
\end{eqnarray}
where 
\begin{eqnarray}
\frac {d\bar\sigma_R^{IR\; \mp}}{dQ^2}=\mp \frac \alpha \pi (\delta_S+\bar\delta_H)\frac {d\sigma_{B}}{dQ^2}
\label{irr}
\end{eqnarray}
does not depend on the separation parameter $\bar v$ introduced in Eq.~(\ref{dsh}). 

After integration of the extracted infrared term over $\phi_k$ and $\tau$ keeping $v$-dependence
only in the denominator, the finite part of hard photon emission reads:
\begin{eqnarray}
&\displaystyle
\frac{d\sigma_R^{F\; \mp}}{dQ^2}=\mp\frac{\alpha ^3 }{2Q^2\lambda_S}
\int\limits_0^{v_{cut}}dv
\Biggl[
\int\limits_{\tau^q_{min}}^{\tau^q_{max}}\frac {d\tau}{(1+\tau)t}
\nonumber\\
&\displaystyle
\times
\sum_{i,j,k=1}^2 
\sum_{l=l^d_{ijk}}^{l^u_{2jk}} 
\frac{\theta_{2jk}^l(Q^2,\tau)}{R^{2-l}}s
F_i(0)
F_j(Q^2)
F_k(t)
\nonumber\\
&\displaystyle
-4\frac {J_H(S,X,0)}{vQ^2}\sum_{i,j=1,2}\theta^B_{ij}F_i(Q^2)F_j(Q^2)\Biggr].
\end{eqnarray}

The sum of Eq.~(\ref{irr}) and Eq.~(\ref{irb}) 
\begin{eqnarray}
\frac{d\sigma^{\mp}_{IR}}{dQ^2}+\frac{d\sigma^{\mp}_{box\; IR}}{dQ^2}
=
\mp \frac \alpha \pi 
\delta_{VR}\frac{d\sigma_{B}}{dQ^2}
\end{eqnarray}
is infrared free since 
\begin{eqnarray}
\delta_{VR}
&=&{\hat \delta}^{IR}_{2\gamma}+\delta_S+
\bar\delta_H
=
2(XL_X-SL_S)\log\frac {v_{cut}}{mM}
\nonumber\\
&&
+\delta _{2\gamma}
+\frac S4 \sqrt{\lambda_S}L_S^2
-\frac X4 \sqrt{\lambda_X}L_X^2
\nonumber\\[1mm]
&&
+\frac S{\sqrt{\lambda_S}}{\rm Li}_2\biggl(\frac {2\sqrt{\lambda_S}}{S+\sqrt{\lambda_S}}\biggr)
\nonumber\\
&&
-\frac X{\sqrt{\lambda_X}}{\rm Li}_2\biggl(\frac {2\sqrt{\lambda_X}}{X+\sqrt{\lambda_X}}\biggr)
\nonumber\\[2mm]
&&
+S_\phi(k_2,p_1,p_2)
-S_\phi(k_1,p_1,p_2)
\end{eqnarray}
does not depend on $P_{IR}$.

Finally, the lowest order of the charge-odd contribution
to the elastic lepton-proton scattering has a form: 
\begin{eqnarray}
\frac{d\sigma^{\mp}_{odd}}{dQ^2}=\frac{d\sigma^{\mp}_F}{dQ^2}
\pm \frac \alpha \pi 
\delta_{VR}\frac{d\sigma_{B}}{dQ^2}.
\label{oddq2}
\end{eqnarray}

\subsection{Contribution to $d\sigma/d\cos\theta$}
The phase space for fixed angle $\theta $ reads:
\begin{eqnarray}
d\Gamma_3&=&J_\theta(v)
\frac{vdv d\cos\theta  d\tau d\phi_k}{2^8\pi^4(1+\tau)^2\sqrt{\lambda_S \lambda_v}},
\end{eqnarray}
where $\lambda_v=(Q^2_R(v)+v)^2+4M^2Q^2_R(v)$ and
\begin{eqnarray}
J_\theta(v)&=&\frac {\lambda_S-v S-Q^2_R(v)(S+2M^2)}
{(S+2M^2)^2-\lambda_S\cos^2\theta}
\nonumber\\&\times&
\Biggl(\frac{S+2M^2}{\cos\theta}+M\sqrt{\frac{\lambda_S}{\cal D}}(S-v+2m^2)\Biggr).
\end{eqnarray}
The square of transferred momentum for the radiative process with a fixed scattering angle
is expressed through the inelasticity $v$ as it was obtained in our previous work \cite{lpcth}:
\begin{eqnarray}
Q_R^2(v)&=&\frac 1{(S+2M^2)^2-\lambda_S\cos^2\theta}
\nn&&\times
\biggl[(S+2M^2)(\lambda_S-vS)
-\lambda_S (S-v)\cos^2\theta
\nn&&
-2 M\sqrt{\lambda_S}
\sqrt{{\cal D}}\cos\theta\biggl],
\label{q2v}
\end{eqnarray}
where
\begin{eqnarray}
{\cal D}=M^2(\lambda_S+v(v-2S))-m^2(\lambda_S\sin^2\theta+4vM^2).
\nn
\end{eqnarray}
It should be noticed that $Q_R^2(0)=Q^2$
and $J_\theta(0)=j_\theta$.

After some algebra similar to the previous subsection we find that:
\begin{eqnarray}
\frac{d\sigma^{\mp}_{odd}}{d\cos\theta}=\frac{d\sigma^{\mp}_F}{d\cos\theta}
\pm \frac \alpha \pi 
\delta_{VR}\frac{d\sigma_{B}}{d\cos\theta}.
\label{oddth}
\end{eqnarray}
The finite part reads:
\begin{eqnarray}
&\displaystyle
\frac{d\sigma_F^\mp}{d\cos\theta}=\mp\frac{\alpha ^3 }{2\lambda_S}
\int\limits_0^{v_{cut}}dv
\Biggl[J_\theta(v)
\int\limits_{\tau^\theta_{min}}^{\tau^\theta_{max}}
\frac{d\tau}{tQ^2_R(v)(1+\tau)}
\nonumber\\
&\displaystyle
\times
\sum_{i,j,k=1}^2 
\sum_{l=l^d_{ijk}}^{l^u_{2jk}} 
\frac{\theta_{2jk}^l(Q^2_R(v),\tau)}
{R^{2-l}}
F_i(0)
F_j(Q^2_R(v))
F_k(t)
\nonumber\\
&\displaystyle
-4
\frac {j_\theta}{vQ^4}J_H(S,X,0)\sum_{i,j=1}^2\theta^B_{ij}
F_i(Q^2)
F_j(Q^2)
\Biggr],
\end{eqnarray}
where the range of $\tau $ at fixed $\theta$ is defined by 
\begin{eqnarray}
\tau^{\theta}_{max/min}=\frac{Q^2_R(v)+v\pm\sqrt{\lambda_v}}{2M^2}.
\end{eqnarray}

\section{Numerical results}
\label{na}

\begin{figure*}[hbt]\centering
\vspace*{-6mm}
\includegraphics[width=70mm,height=70mm]{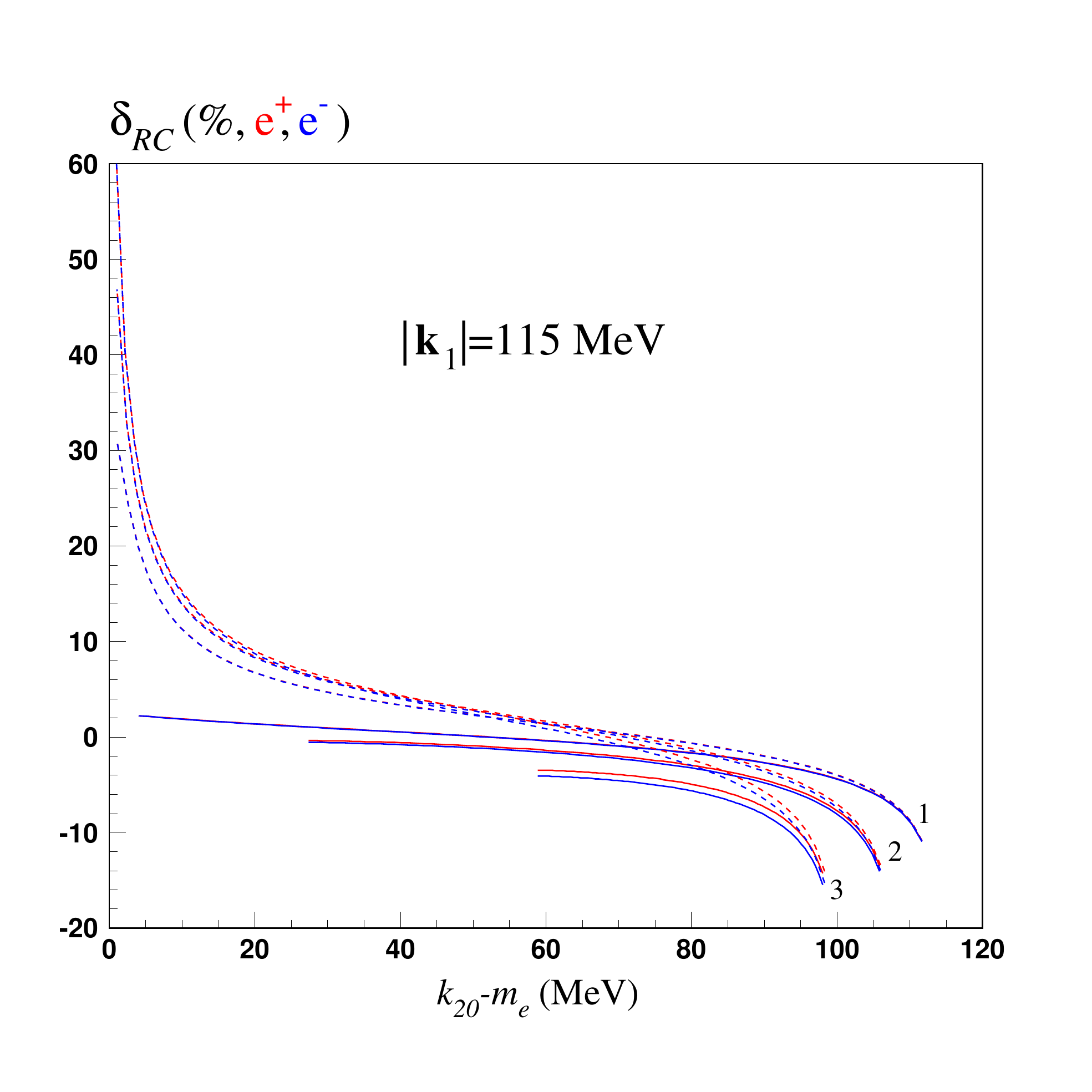}
\hspace*{-6mm}
\includegraphics[width=70mm,height=70mm]{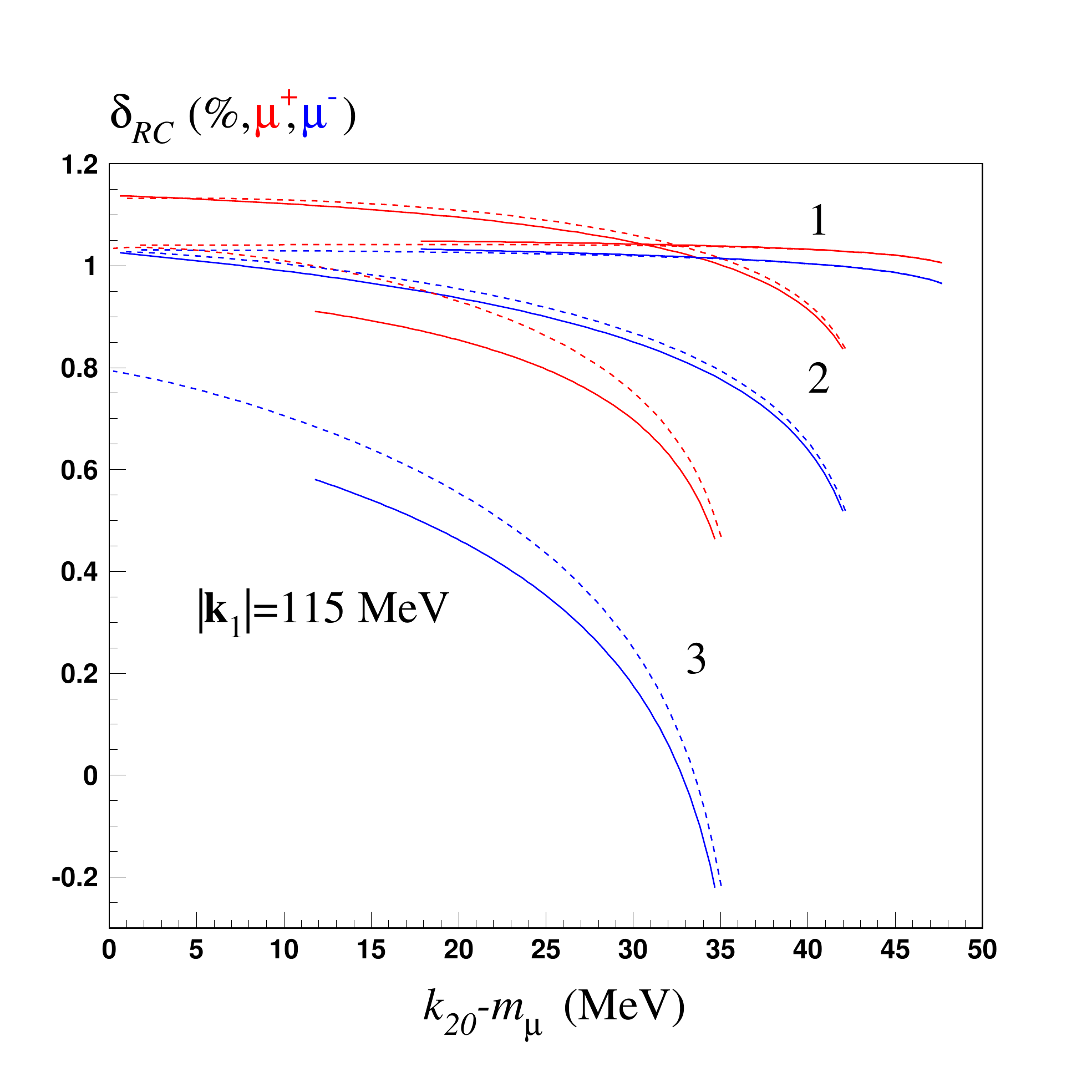}
\\[-9mm]
\includegraphics[width=70mm,height=70mm]{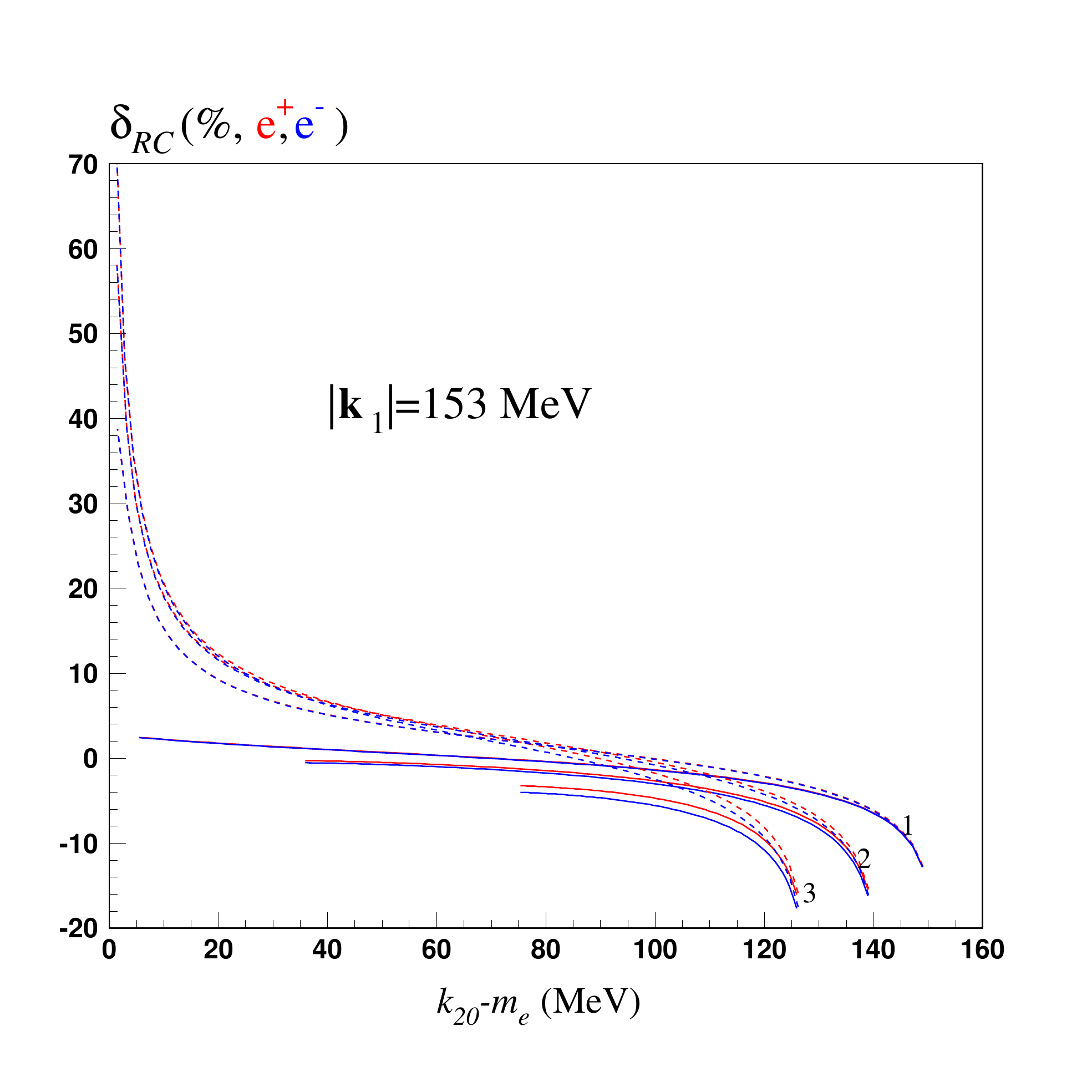}
\hspace*{-6mm}
\includegraphics[width=70mm,height=70mm]{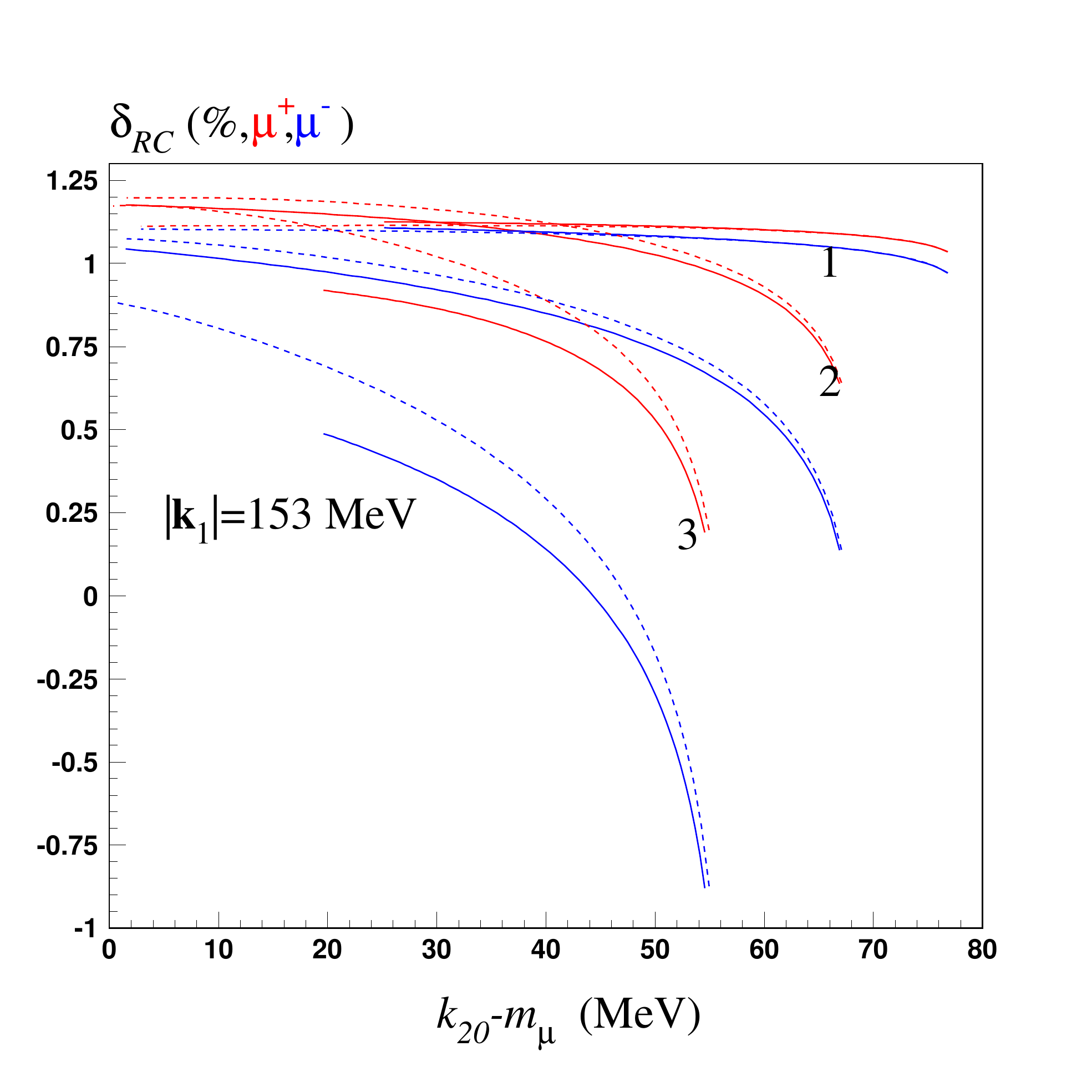}
\\[-9mm]
\includegraphics[width=70mm,height=70mm]{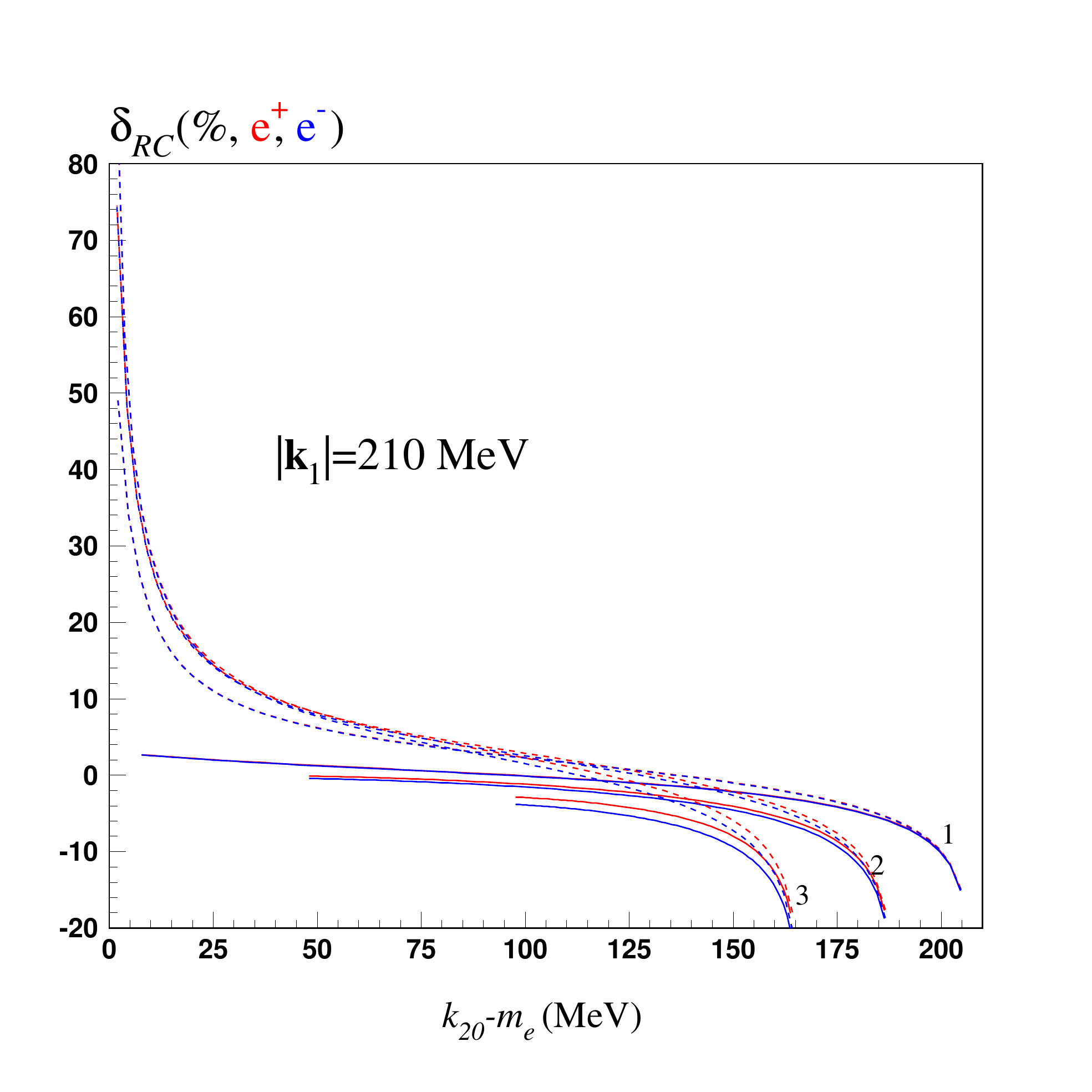}
\hspace*{-6mm}
\includegraphics[width=70mm,height=70mm]{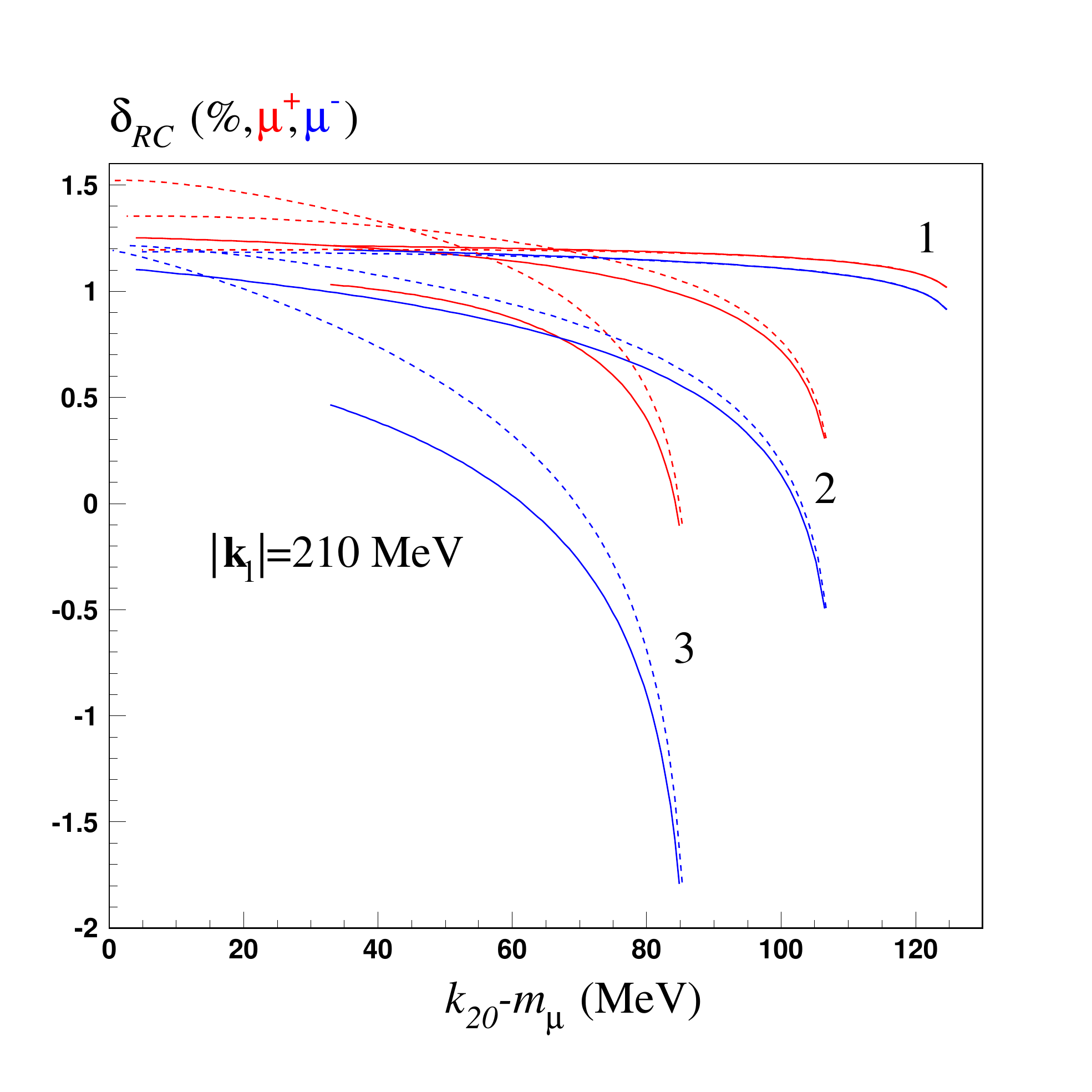}
\\
\caption{Relative RC vs the value of the scattering lepton kinetic energy for elastic $e^\mp p$ and $\mu^\mp p$ scattering, beam momenta is equal to 115 MeV, 153 MeV and 210 MeV for $\theta=20^o$ (1), $60^o$ (2), $100^o$ (3). Blue (red)  lines correspond to lepton (antilepton) scattering while solid (dashed) lines correspond to fixed $Q^2$ ($\cos\theta $). 
}
\vspace*{-3mm}
\label{fig4a}
\end{figure*}

For understanding the distinguish between radiative effects for incoming lepton and antilepton
similar to the work \cite{lpcth} we present a ratio
of RC for particle ($d\sigma^-_{RC}/d\zeta$) and antiparticle ($d\sigma^+_{RC}/d\zeta$) to the Born cross section:
\begin{eqnarray}
\delta^\mp_{RC}=\frac{d\sigma^\mp_{RC}/d\zeta}{d\sigma_{B}/d\zeta}
\label{del}
\end{eqnarray}
both for fixed $Q^2$ ($\zeta=Q^2$) and the scattering angle ($\zeta=\cos\theta$). Here 
the quantities $d\sigma^\mp_{RC}/d\zeta$ contain from the sum of the charge-even RC calculated in \cite{lpcth} and 
$d\sigma^\mp_{odd}/d\zeta$ defined by Eqs.~(\ref{oddq2}) or (\ref{oddth}) in dependence on $\zeta $.

Notice, that for the radiative process the scattering lepton energy $k_{20}$ depends on the inelasticity $v$ in relation to the selected variable. So for fixed $Q^2$ 
\begin{eqnarray}
k_{20}=\frac{X-v}{2M}
\label{k201}
\end{eqnarray}
while for fixed scattering angle
\begin{eqnarray}
k_{20}=\frac{S-Q^2_R(v)-v}{2M}.
\label{k202}
\end{eqnarray}

For the parametrization of the proton form factors the results of Kelly's paper \cite{Kelly} is used.  

From Fig.~\ref{fig4a}, where the relative RC (\ref{del}) as a function of the scattering lepton kinetic energy  are presented for particle (blue curves) and antiparticle (red curves) scattering  at MUSE kinematic conditions \cite{MUSE}, it can be seen that RC are greater for antilepton than for lepton 
and their difference reaches maximum value at large scattering angles and soft photon emission.

\begin{figure*}[hbt]\centering
\vspace*{-6mm}
\includegraphics[width=70mm,height=70mm]{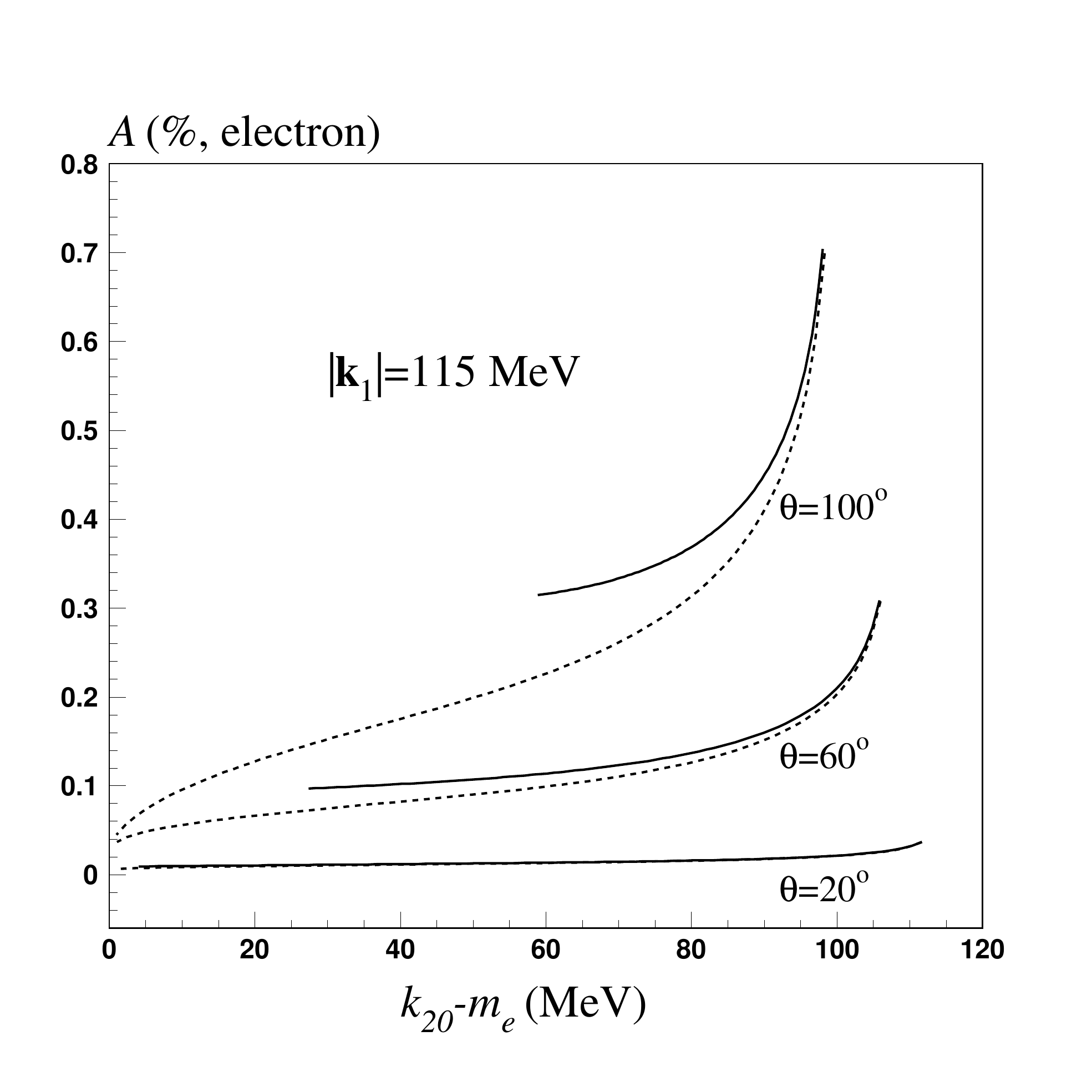}
\hspace*{-6mm}
\includegraphics[width=70mm,height=70mm]{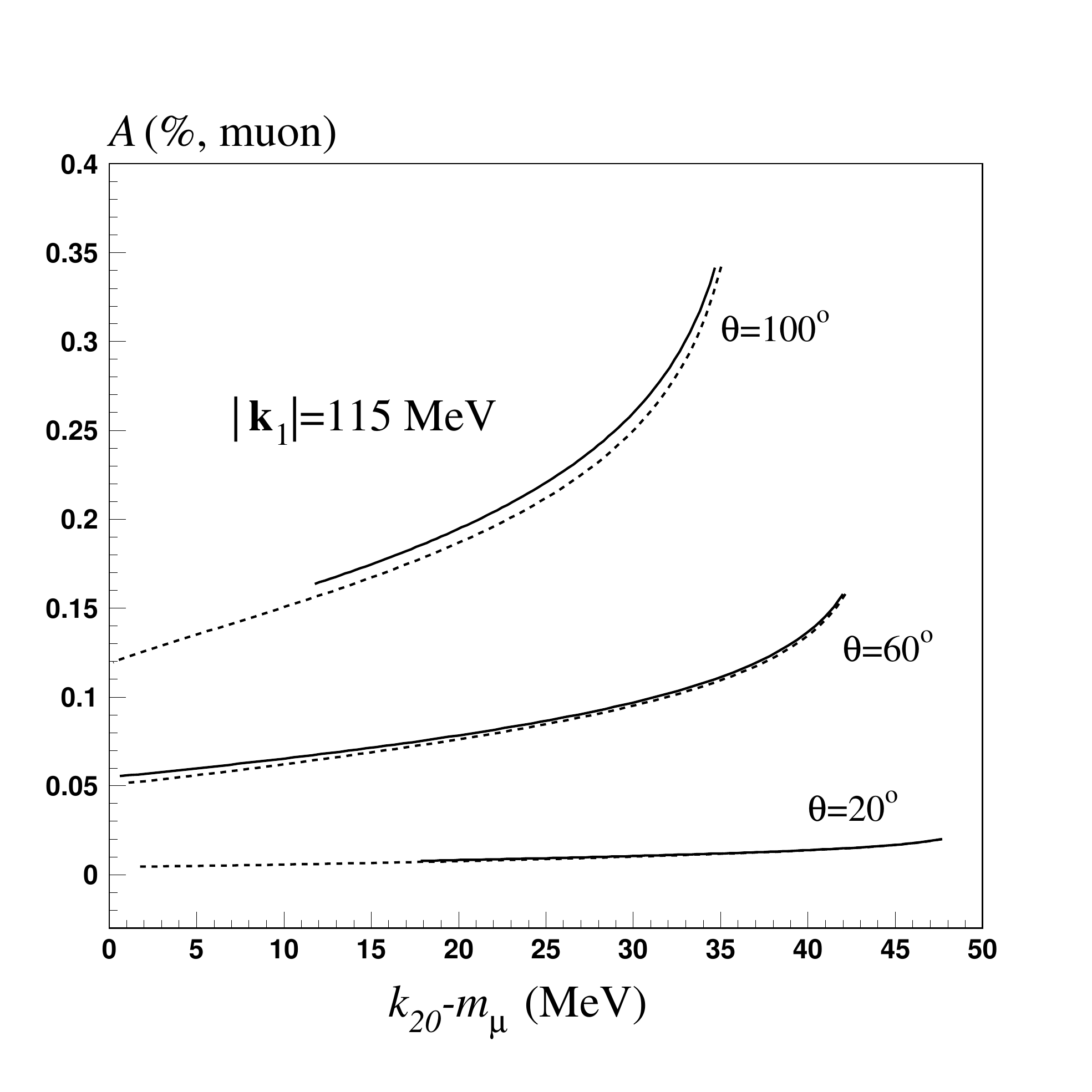}
\\[-9mm]
\includegraphics[width=70mm,height=70mm]{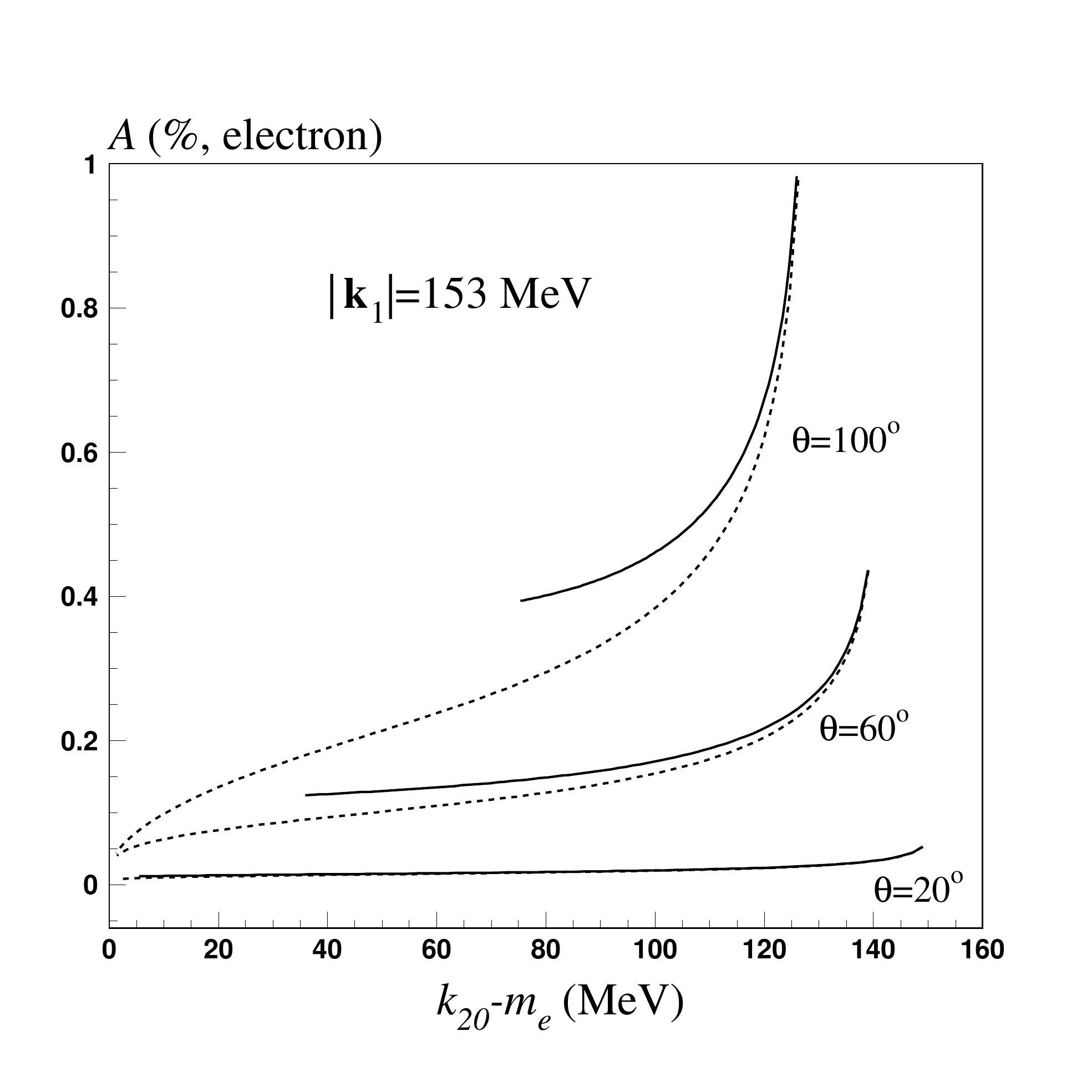}
\hspace*{-6mm}
\includegraphics[width=70mm,height=70mm]{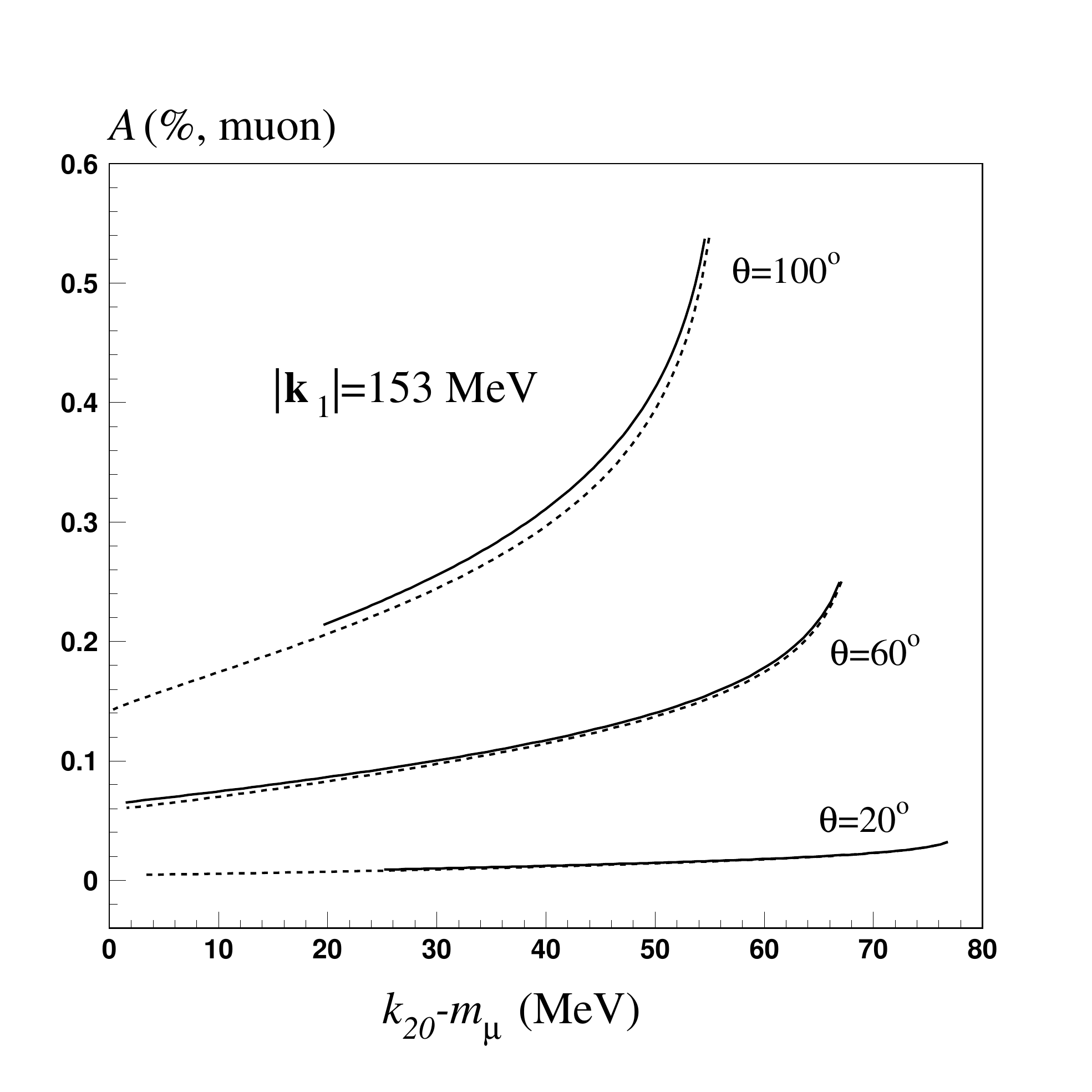}
\\[-9mm]
\includegraphics[width=70mm,height=70mm]{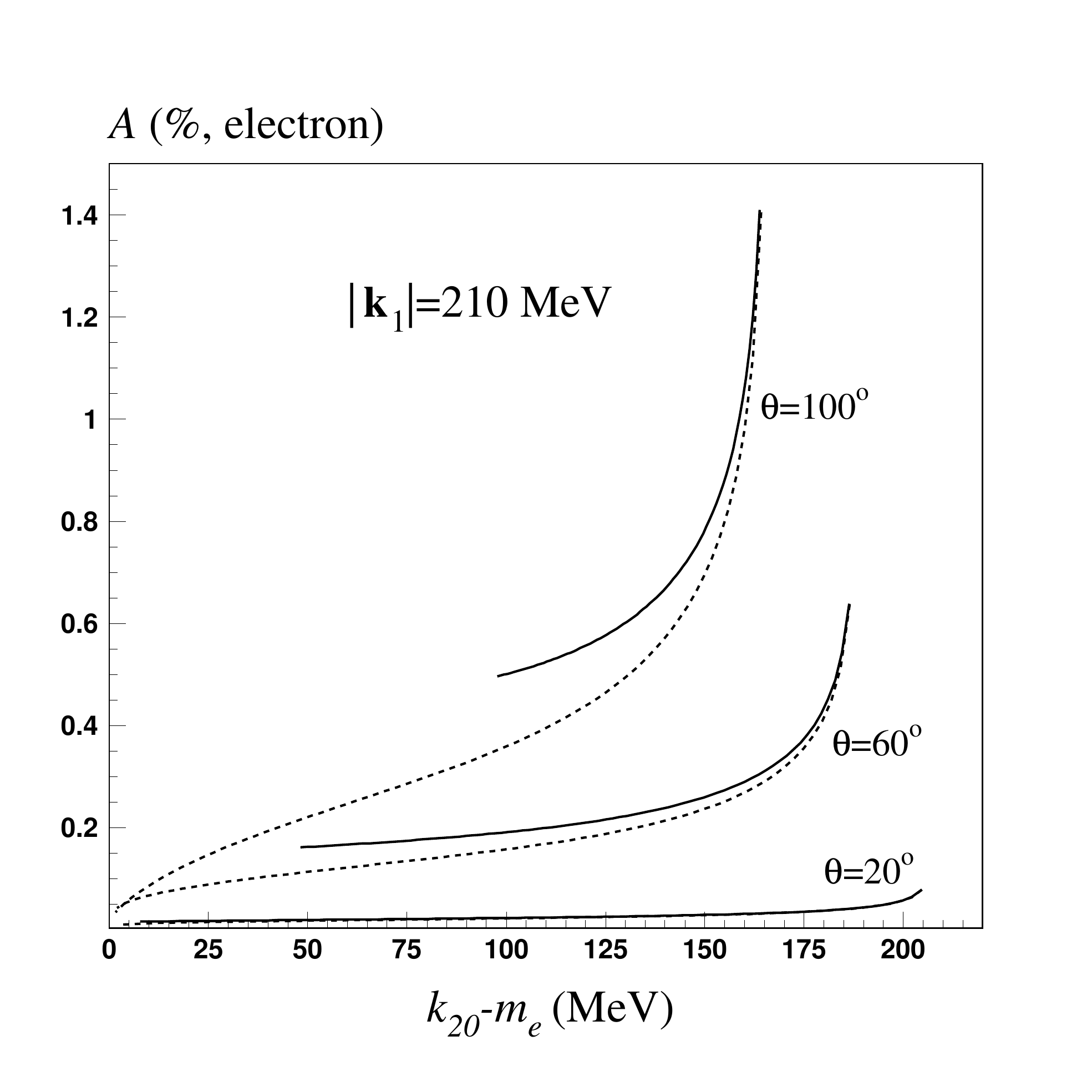}
\hspace*{-6mm}
\includegraphics[width=70mm,height=70mm]{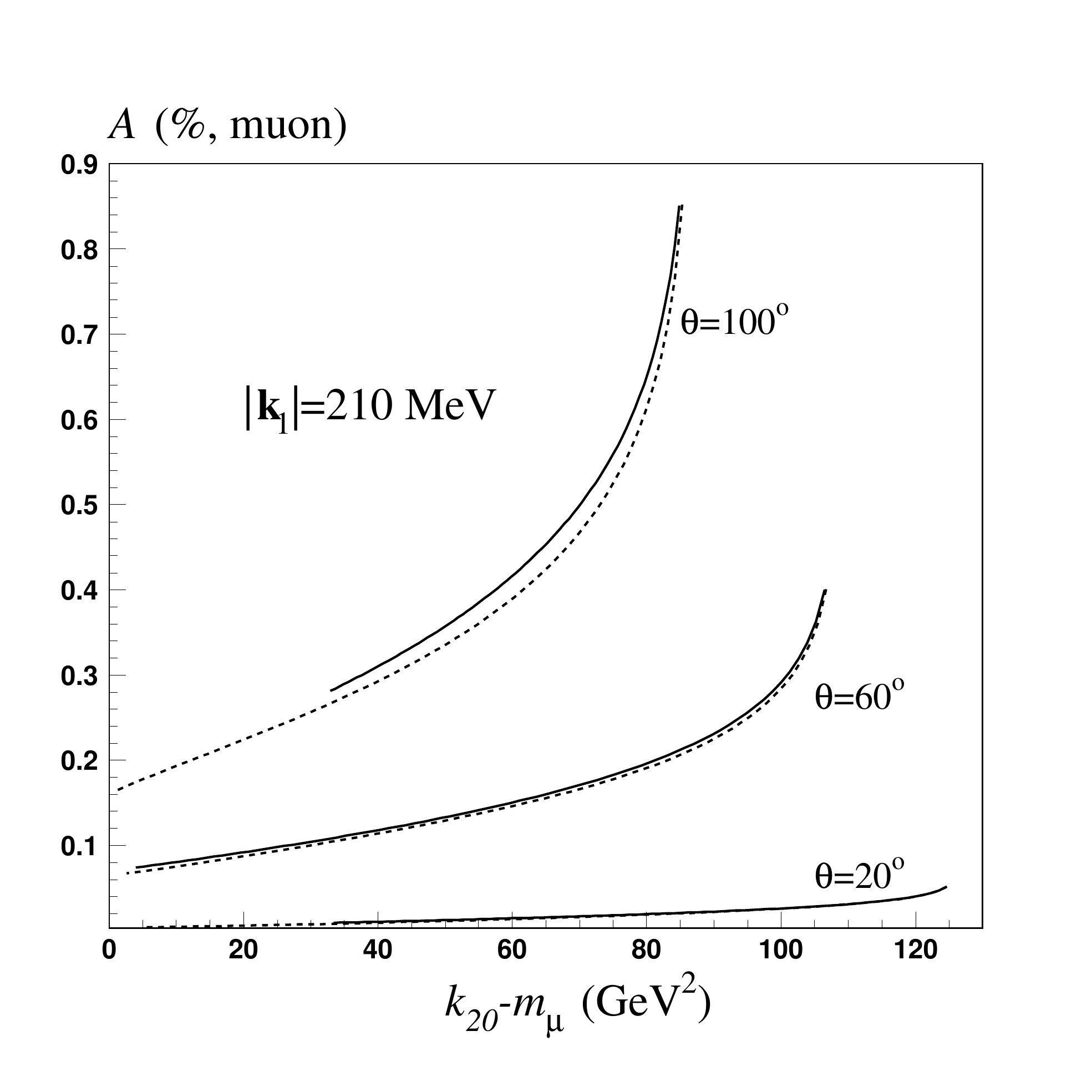}
\\
\caption{Charge asymmetry defined in (\ref{casy}) vs value of the scattering particle kinetic energy
for elastic $e^\mp p$ and $\mu^\mp p$ scattering with the beam momenta 115 MeV, 153 MeV and 210 MeV.
The solid (dashed) correspond fixed $Q^2$ ($\cos\theta $). 
}
\label{fig4}
\end{figure*}

Using above presented results the positively defined charge asymmetry can be written as 
\begin{eqnarray}
A_{\zeta}=\frac{d\sigma^{+}_{odd}/d\zeta-d\sigma^{-}_{odd}/d\zeta}{d\sigma_{even}/d\zeta},
\label{casy}
\end{eqnarray}
where the even cross section include Born contribution and RC to the lepton line that was found in \cite{lpcth}.

The dependence of this asymmetry on the scattering lepton kinetic energy for different lepton beams presented in Fig.~\ref{fig4} at MUSE kinematic conditions \cite{MUSE} at different scattering angles. 
As can be seen, the asymmetry reaches its maximum value for the lightest lepton with the highest of the initial and scattering lepton energies at the maximum $\theta $. At the soft photon region its behavior is almost identical both for fixed $Q^2$ and for fixed $\theta$ and has a maximum difference for these two observables at low scattering electron energy and high $\theta $. The asymmetry value for fixed $\theta $ is always less than its value for fixed $Q^2$.

Another rather interesting quantity is the dependence of $e^+p/e^-p$ cross section ratio 
\begin{eqnarray}
R=\frac{d\sigma_{even}/d\xi+d\sigma^{+}_{odd}/d\xi}{d\sigma_{even}/d\xi +d\sigma^{-}_{odd}/d\xi }
\label{rat}
\end{eqnarray}
on the virtual photon polarization $\varepsilon$. 
According to \cite{Preedom},
$\varepsilon$ beyond the ultrarelativistic approximation has a form
\begin{eqnarray}
\varepsilon=\Biggl(1+2(1+\tau_p )\frac{(Q^2-2m^2)}{4k_{10}k_{20}-Q^2}\Biggr)^{-1}.
\label{eps}
\end{eqnarray}
Since for radiative processes the scattered lepton energy depends on the inelasticity as shown in Eqs.~(\ref{k201}, \ref{k202}), setting any cut on $v$ leads to $\varepsilon$ restriction.

\begin{figure}[t]
\vspace*{-5mm}
\hspace*{2mm}
\includegraphics[width=75mm,height=75mm]{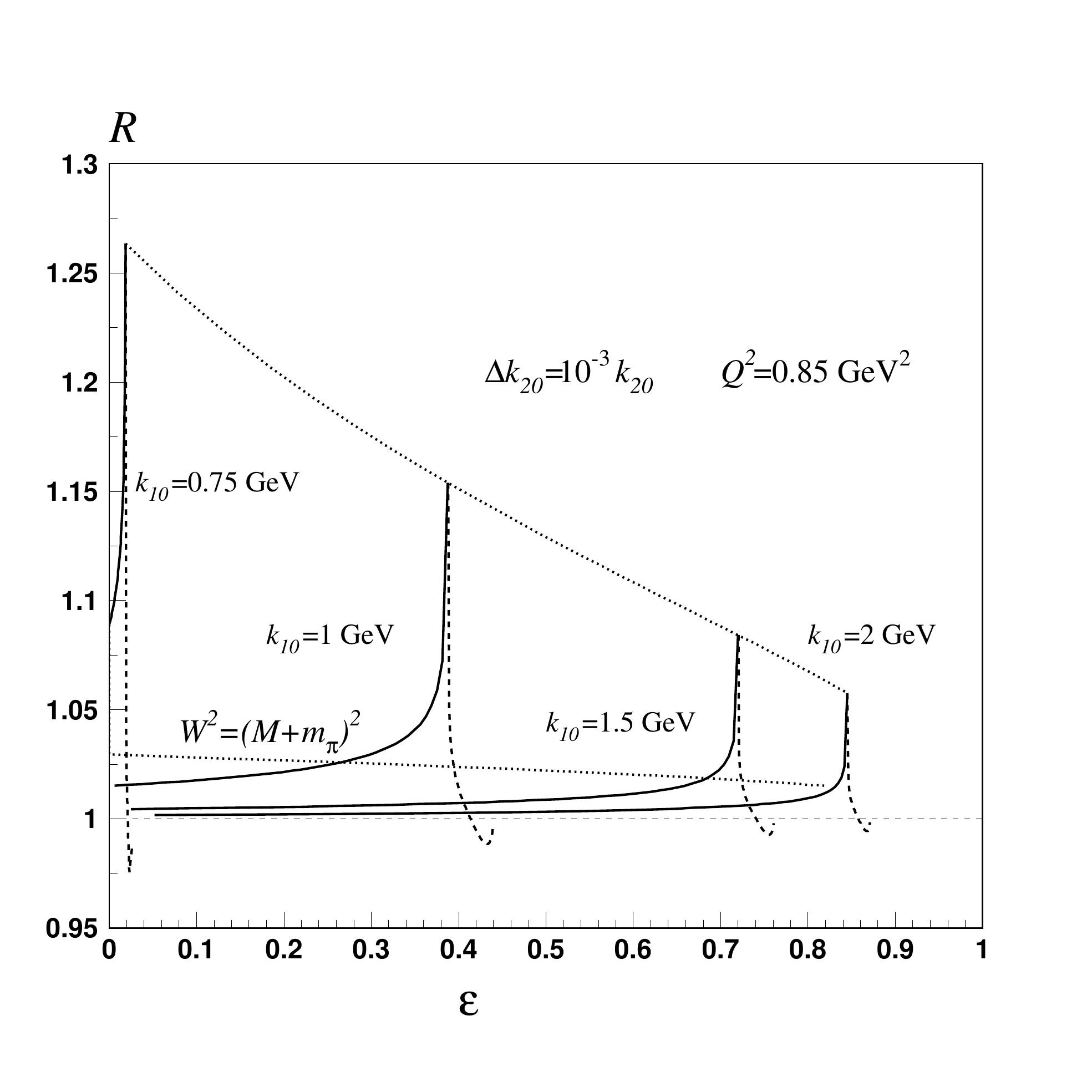}
\\[-11mm]
\hspace*{2mm}
\includegraphics[width=75mm,height=75mm]{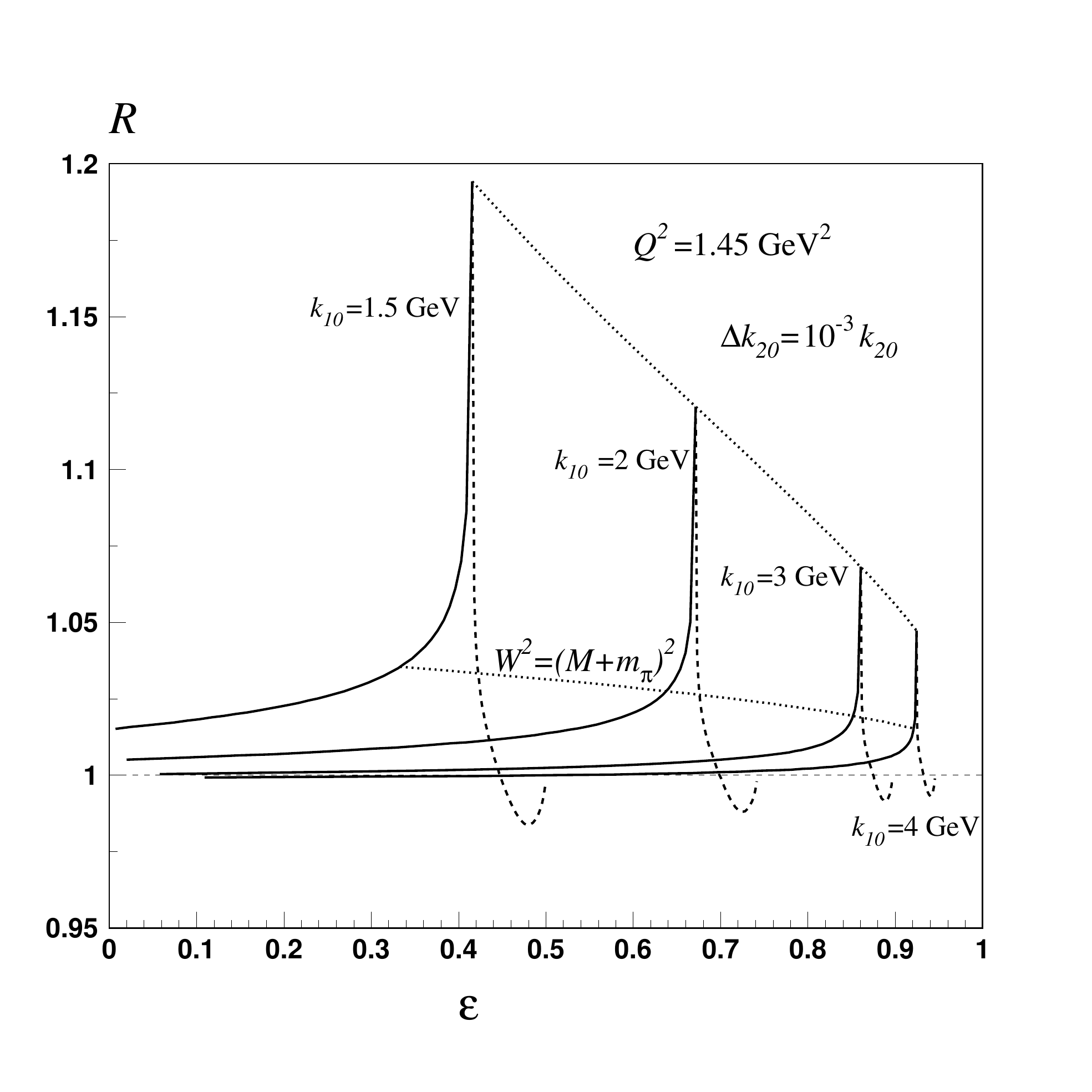}
\\[-4mm]
\caption{
Ratio of $e^+p/e^-p$ cross sections defined in (\ref{rat}) as a function of $\varepsilon$ at $Q^2=0.85$ GeV$^2$ and $Q^2=1.45$ GeV$^2$  for
the different electron beam energies $k_{10}$ defined in the rest
frame (${\bf p}_1{\bf = 0}$).
The solid (dashed) line corresponds to fixed $Q^2$ ($\cos\theta $).  An upper dotted line represents soft photon emission. The lowest dotted lines correspond to the pion
production threshold.
}
\label{fig5}
\vspace*{-2mm}
\end{figure}

The obtained dependence at JLab kinematic conditions is shown in Fig.~\ref{fig5}. For soft photon emission the cross section ratio reaches its maximum value, and, similar to the work \cite{Jlab}, increases with decreasing $\varepsilon$. For hard photon emission the ratio decreases with decreasing $\varepsilon$ for fixed $Q^2$ and increasing $\varepsilon$ for fixed scattering angle.

\section{Conclusion}

The contribution of hard photon emission to the charge asymmetry in lepton- and antilepton-proton scattering 
was calculated and numerically compared for fixed $Q^2$ and fixed scattering angle beyond the ultrarelativistic limit keeping lepton mass during 
the whole process of calculation.

During the calculation only two assumptions are used: I) We did not consider any excited states in the intermediated proton in two-photon exchange,
that allows to use a standard fermionic propagator; II) The on-shell proton vertex with the Dirac and Pauli form factors 
 is applicable within off-shell region.

Two-photon exchange was analytically calculated
in soft photon approximation using Tsai approach and  dimensional regularization of the infrared divergence.
The infrared divergence from real photon emission was treated by the Bardin-Shumeiko technique that allow to rid dependence of the final results on any artificial parameter for soft scale separation.

The numerical results presented for kinematic conditions of  MUSE experiment at PSI and  Jefferson Lab
measurement shown that the asymmetry reaches its maximum value at the soft photon emission and significantly drops for hard real photon emission. 

We believe that the results presented here and in our previous work \cite{lpcth}, such as the derivation of the essential formulae, the sets of the explicit expressions are quite transparent and useful for studying both the treatment of hard photon emission and soft photon extraction from two-photon exchange. Moreover, the implementation of the obtained results into Monte-Carlo generator ELRADGEN \cite{ELRADGEN1,ELRADGEN2} allows to develop an alternative
to the generator ESEPP \cite{ESEPP} for the simulation of hard photon emission in lepton- and antilepton-proton scattering in rather wide kinematical region from low $Q^2$ - relevant to ``proton radius puzzle” - to high $Q^2$ where the form factor ratio problem was observed.

\section*{Acknowledgements}

Work of AA was supported by National Science Foundation under grant No. PHY-1812343.


\appendix
\section{Calculation of the three-point loop integrals.}
\label{tpli}
Here we present the details of calculation of the three-point loop integrals defined by Eqs.~(\ref{kab})

Using Feynman parametrization we find that 
\begin{eqnarray}
K_{IR}(a,b)=-\frac {4ab}{i\pi^2}\int\limits_0^1dy 
\int\limits_0^1dx 
\int\frac{x(2\pi \mu)^{4-n}d^nl}{(l^2-x^2c_y^2(a,b))^3}\;\;
\end{eqnarray}
with $c_y(a,b)=a y+b(1-y)$.

The integration over $l$ in $n$-dimensional space gives:
\begin{eqnarray}
K_{IR}(a,b)&=&2ab\frac{\Gamma\left(3-n/2\right)}{(2\sqrt{\pi}\mu)^{n-4}}
\int\limits_0^1dy 
\int\limits_0^1dx x^{n-5}c^{n-6}_y(a,b).
\nonumber\\
\label{jd2}
\end{eqnarray}
After integration over $x$, and the expansion of the obtained expressions
into the Laurent series around $n=4$ result in: 
\begin{eqnarray}
K_{IR}(a,b)&=&ab 
\int\limits_0^1 
\frac{dy}{c^2_y(a,b)}\Biggl[
\displaystyle
2P_{IR}
+
\log \frac{c^2_y}{\mu^2}
\Biggr]
\nn
&=&2ab\int\limits_0^1 
dy{\cal K}_y(a,b),
\label{kab3}
\end{eqnarray}
where the term representing the infrared divergence in the dimensional regularization reads
\begin{eqnarray}
P_{IR}=\frac 1{n-4}+\frac 12\gamma _E+\log\frac 1{2\sqrt{\pi}}.
\label{pir}
\end{eqnarray}
After substitution into (\ref{kab3}) photon mass regularization $P_{IR}=\log \mu/\lambda $ we immediately find that our definition of $K_{IR}(a,b)$ is equal to $K(p_i,p_j)$ defined by Eq.~(I.5) of \cite{Tsai1961}.

Taking into account that $c^2_y(k_1,-p_1)=c^2_y(k_2,-p_2)\\=\zeta_d(y)$ and $c^2_y(k_1,p_2)=c^2_y(k_2,p_1)=\zeta_x(y)$, where
\begin{eqnarray}
\zeta_d(y)
&=&y (m^2 y-S (1-y))+M^2 (y-1)^2,
\nn
\zeta_x(y)
&=&y (m^2 y+X (1-y))+M^2 (y-1)^2
\end{eqnarray}
we will consider only two integrals, namely $K_d=K_{IR}(k_1,-p_1)=K_{IR}(k_2,-p_2)$ and  $K_s=K_{IR}(k_1,p_2)=K_{IR}(k_2,p_1)$, that defined in the following way:
\begin{eqnarray}
K_d&=&-\frac S2 
\int\limits_0^1 
\frac{dy}{\zeta_d(y)}\Biggl[
\displaystyle
2P_{IR}
+
\log \frac{\zeta_d(y)}{\mu^2}
\Biggr]
=\int\limits_0^1 
dy{\cal K}_d(y),
\nn
K_x&=&\frac X2 
\int\limits_0^1 
\frac{dy}{\zeta_x(y)}\Biggl[
\displaystyle
2P_{IR}
+
\log \frac{\zeta_x(y)}{\mu^2}
\Biggr]
=\int\limits_0^1 
dy{\cal K}_x(y).
\nn
\label{kdx}
\end{eqnarray}

\includegraphics[width=7cm,height=3.5cm]{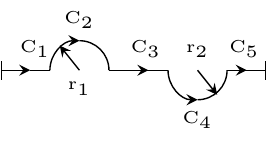}

\noindent
{\small {\bf Fig. 7 }Path integration over $y$ for $K_d$.}
\vspace*{3mm}
The expression $\zeta_{d,x}(y)$ can be presented in the following way:
\begin{eqnarray}
\zeta_{d,x}(y)=\frac{M^2}{y_1^{d,x}y_2^{d,x}}(y-y_1^{d,x})(y-y_2^{d,x}),
\label{zdx}
\end{eqnarray}
where
\begin{eqnarray}
y_1^d=\frac{2M^2}{2M^2+S+\sls},\;
y_2^d=\frac{2M^2}{2M^2+S-\sls},\;\;\;\;
\nonumber\\
y_1^x=\frac{2M^2}{2M^2-X+\slx},\;
y_2^x=\frac{2M^2}{2M^2-X-\slx}.\;\;\;\;
\end{eqnarray}

It should be noted that for all $S>2mM$ the quantities $y_{1,2}^d$ belong
to the region of the integration:
$1>y_2^d>y_1^d>0$. As a result, according to Eq.~(\ref{zdx}) the function
$\zeta_{d}(y)$ is positive for the two segments of the integration, namely
$0<y<y_1^d$ and $y_2^d<y<1$, and negative between them. Moreover
at the points $y=y_{1,2}^d$ the integral over $y$ in
$K_d$ defined by Eq. (\ref{kdx}) diverges.

To perform the integration over $y$ the method suggested by J.~Kahane in Ref.~\cite{Kahane} is used. For this purpose the integration region is
broken up into five segments as it is shown in Fig.~7.
The contours $C_2$ and $C_4$ are chosen such that  $\zeta_{d}(y)$
has negative imaginary parts.

The integration for the regions $C_1$, $C_3$ and $C_5$
can be expressed via  
Spence's dilogarithm
\begin{eqnarray}
{\rm Li }_2(x)=-\int\limits_0^x\frac {\log |1-y|}ydy
\end{eqnarray}
in a following way

\vspace*{12mm}
\begin{strip}
\rule[-1ex]{\columnwidth}{1pt}\rule[-1ex]{1pt}{1.5ex}
\begin{eqnarray}
K_{d}^{C_1}&=&
\lim_{\delta_1\to 0}\int\limits_0^{y_1^d-\delta_1} 
dy
{\cal K}_d(y)
=\frac S{4\sls}
\Biggl[
2\biggl(2P_{IR}+\log\frac\sls{\mu ^2}\biggr)
\log \frac{\delta_1(S+2M^2+\sls)^2}{4M^2\sls}
+
\log^2\delta_1
\nonumber\\&&
+2
\log\frac{S+2M^2+\sls}{2\sls}
\log\frac{\sls(S+2M^2+\sls)}{2(S+M^2+m^2)^2}
-\log ^2\frac {M^2}{\sls}
-4
{\rm Li}_2\frac{S+2M^2+\sls}{2\sls}
+\frac 23\pi^2
\Biggr],
\nonumber\\
K_{d}^{C_3}&=&
\lim_{\delta_{1,2}\to 0}\int\limits_{y_1^d+\delta_1}^{y_2^d-\delta_2} 
dy{\cal K}_d(y)
=-\frac S{4\sls}
\Biggl[
2
\biggl(2P_{IR}+\log\frac\sls{\mu ^2}\biggr)
\log\frac {\delta_1\delta_2(S+M^2+m^2)^2}{\lambda_S}
+
\log^2\delta_1
+
\log^2\delta_2
\nonumber\\&&
-2\log^2 \frac {S+M^2+m^2}{\sls}
+\frac 23\pi^2
+2i\pi \log \frac{\delta_1\delta_2(S+M^2+m^2)^2}{\lambda_S}
\Biggr],
\nonumber\\
K_{d}^{C_5}&=&
\lim_{\delta_2\to 0}\int\limits^1_{y_2^d+\delta_2} 
dy{\cal K}_d(y)
=\frac S{4\sls}
\Biggl[
2\biggl(2P_{IR}+\log\frac\sls{\mu ^2}\biggr)
\log \frac{\delta_2(S+2m^2+\sls)^2}{4m^2\sls}
+
\log^2\delta_2
\nonumber\\&&
+2
\log\frac{S+2m^2+\sls}{2\sls}
\log\frac{2m^4}{\sls(S+2m^2+\sls)}
-\log ^2\frac {m^2}{\sls}
+4
{\rm Li}_2\frac{\sls-S-2m^2}{2\sls}
\Biggr].
\end{eqnarray}

The integration along $C_2$ and $C_4$ is done by replacing 
$y\to r_1=y_1^d-\delta_1 \exp(-i\theta)$ and $y\to r_2=y_2^d-\delta_1 \exp(i\theta)$
respectively:
\begin{eqnarray}
K_{d}^{C_2}&=&
\lim_{\delta_1\to 0}\int\limits_0^{\pi} 
d\theta \frac{dr_1}{d\theta}{\cal K}_d(r_1)
=
-\frac S{4\sls}
\Biggl[\pi^2+2i\pi
\biggl(2P_{IR}+\log\frac{\delta_1\sls}{\mu ^2}\biggr)
\Biggr],
\nonumber\\
K_{d}^{C_4}&=&
\lim_{\delta_2\to 0}\int\limits_0^{\pi} 
d\theta \frac{dr_2}{d\theta}{\cal K}_d(r_2)
=-\frac S{4\sls}
\Biggl[\pi^2+2i\pi
\biggl(2P_{IR}+\log\frac{\delta_2\sls}{\mu ^2}\biggr)
\Biggr]
.
\end{eqnarray}

Summing up the integral over all five segments we obtained that: 
\begin{eqnarray}
K_{d}&=&\sum_{i=1}^5K_{d}^{C_i}
=\frac S{4\sls}\Biggl[
\biggl(4P_{IR}+4\log \frac{m}{\mu}-\log\frac {S+\sls}{S-\sls}\biggr)\log\frac {S+\sls}{S-\sls}
+4{\rm Li}_2\frac{\sls-S-2m^2}{2\sls}
\nonumber\\&&   
-4
{\rm Li}_2\frac{S+2M^2+\sls}{2\sls}
+2\log\frac{S-\sls}{2M^2}\log\frac{(S-\sls)(S+2M^2-\sls)^2}{8M^2\lambda_S}
-2\pi^2
\nonumber\\&&
-4i\pi
\biggl(2P_{IR}+\log\frac {\delta _1\delta _2(S+m^2+M^2)}{\mu^2}\biggr)
\Biggr].
\end{eqnarray}

As opposed to $\zeta_{d}(y)$ in the region $0<y<1$ the other function $\zeta_{x}(y)$
is always positive since $y_2^x>y_1^x>1$ for $2mM<X<M^2+m^2$ and $y_1^x>1>0>y_2^x$
for $X>M^2+m^2$. For both these situations we have:
\begin{eqnarray}
K_{x}&=&
\frac X{4\slx}\Biggl[
\biggl(4P_{IR}+4\log \frac{m}{\mu}+\log\frac {X+\slx}{X-\slx}\biggr)\log\frac {X+\slx}{X-\slx}
+4
{\rm Li}_2\frac{2M^2+\slx-X}{2\slx}
\nonumber\\&&
-4{\rm Li}_2\frac{X+\slx-2m^2}{2\slx}
-2\log\frac{X+\slx}{2M^2}\log\frac{(X+\slx)(X-2M^2+\slx)^2}{8M^2\lambda_X}
\Biggr].
\end{eqnarray}

\section{Explicit expression for $\theta _{ijk}^l(Q^2,\tau )$}
\label{thex}
The quantities $\theta _{ijk}^l(Q^2,\tau )$ read:

\begin{eqnarray}
\theta^1_{111}&=&4(S^2+X^2-2 Q^2 (m^2+M^2))F_{IR},
\nonumber\\[1mm]
\theta^2_{111}&=&4S_p((Q^2-2M^2)F+2m^2F_d-(Q^2+\tau S )F_{1+} )
+\tau Q^2(4(m^2+M^2)-Q^2)(\tau F_d+F_{1+})
\nonumber\\[1mm]&&
+2((1+\tau) (5 Q^2-6 S)+2(4-\tau)m^2-2\tau M^2)F_{IR}
+2(2 S^2+(Q^2+\tau S+4m^2)Q^2)F_{1+}
\nonumber\\[1mm]&&
+\frac 4{1+\tau }(S_p((Q^2+2M^2)F-2m^2F_d)+(2m^2Q^2-S X)F_{1+}),
\nonumber\\[1mm]
\theta^3_{111}&=&
2 \tau (2(S_p+X)F+(\tau^2 (M^2+m^2)-(1+\tau)(\tau Q^2+4m^2))F_d
+(\tau(4S-Q^2+m^2+M^2)
\nonumber\\[1mm]&&
+2S+3Q^2)F_{1+})+2(4+8\tau+5\tau^2)F_{IR}
+4((Q^2+2 m^2)F_{1+}+(4M^2-Q^2)F_{z_1}-(S+Q^2)F)
\nonumber\\[1mm]&&
+\frac 4{1+\tau} (XF-Q^2F_{z_1}),
\nonumber\\[1mm]
\theta^4_{111}&=&-\tau (4(1+\tau)F+4F_{z_1}+(2+4\tau+3\tau^2)(\tau F_d+F_{1+})),
\nonumber\\[1mm]
\theta^1_{112}&=&4Q^2(Q^2-2 m^2)F_{IR},
\nonumber\\[1mm]
\theta^2_{112}&=&2Q^2\tau \biggl[
\frac{S_pF-m^2F_{1+}}{1+\tau}
-\tau(Q^2-2 m^2)F_d-(Q^2-m^2 )F_{1+}
\biggr]
+\tau (5 Q^2-8 m^2)F_{IR},
\nonumber\\[1mm]
\theta^3_{112}&=&
2\tau(S+S_p)F+2\tau^2F_{IR}
+\frac \tau 2\biggl[\tau^2 (8 m^2-5 Q^2)F_d-(5\tau Q^2+4(1-\tau)m^2)F_{1+}\biggr]
-8Q^2F_{z_1}
\nonumber\\[1mm]&&
-\frac {2\tau}{1+\tau}((S_p+X)F-Q^2F_{z_1}-m^2F_{1+}),
\nonumber\\[1mm]
\theta^4_{112}&=&-6\tau F_{z_1}-\tau^3(\tau F_d+F_{1+}),
\nonumber\\[1mm]
\theta^1_{121}&=&4Q^2 (Q^2-2 m^2)F_{IR}, 
\nonumber\\[1mm]
\theta^2_{121}&=&\tau Q^2\biggl[2\tau (2m^2-Q^2)F_d+3F_{IR}+2(3m^2-Q^2)F_{1+}
+\frac{2\tau}{1+\tau }(m^2F_{1+}-S_pF)\biggr],
\nonumber\\[1mm]
\theta^3_{121}&=&2(1+\tau)Q^2F+\tau^2(F_{IR}-\frac 32Q^2(\tau F_d+F_{1+}))-\frac{2Q^2}{1+\tau}(F+2F_{z_1}),
\nonumber\\[1mm]
\theta^4_{121}&=&-\frac{\tau^3}2(\tau F_d+F_{1+}),
\nonumber\\[1mm]
\theta^1_{122}&=&\frac{2Q^2}{M^2} (SX+M^2(Q^2-4m^2))F_{IR},
\nonumber\\[1mm]
\theta^2_{122}&=&\frac 1{4M^2}\Biggl[
Q^2\biggl(8Q^2S_pF+\tau^2(Q^2(Q^2-4M^2)+16m^2M^2)F_d
+(16m^2(Q^2+\tau M^2)
\nonumber\\[1mm]&&
+2Q^2((3\tau-2)S-2\tau M^2)+(6+\tau)Q^4-8 \tau S^2)F_{1+}
-\frac{4\tau}{1+\tau}(S_p((Q^2+2M^2)F-2m^2F_d)
\nonumber\\[1mm]&&
+(2m^2Q^2-SX)F_{1+})\biggr)
+2(8m^2(Q^2-\tau M^2)+2Q^2(2\tau M^2-(3+4\tau)S)+(5+3\tau)Q^4+2\tau S^2)F_{IR}
\Biggr],
\nonumber\\[1mm]
\theta^3_{122}&=&\frac 1{8M^2}\Biggl[
4Q^2((6\tau S+(1+\tau )(2M^2-5Q^2)-2X)F+4M^2F_{z_1})
+(4(Q^2+S)(2Q^2-S)
\nonumber\\[1mm]&&
+16m^2(M^2+3Q^2)
+2\tau(8m^2(Q^2-M^2)+9Q^4+2S^2)
+\tau^2(Q^4-8M^2Q^2+22Q^2S-8S^2))F_{1+}
\nonumber\\[1mm]&&
+\tau(\tau Q^2(\tau (Q^2-8M^2)-4Q^2)+16m^2(\tau^2M^2-(1+\tau)Q^2))F_d
+2((1+\tau)(16m^2+\tau(11Q^2-6S))
\nonumber\\[1mm]&&
+2\tau(4Q^2+\tau M^2)+6Q^2) F_{IR}
+\frac 4{1+\tau}((SX-4M^2m^2)F_{1+}-Q^2(S_p+2M^2)F-4M^2Q^2F_{z_1})\Biggr],
\nonumber\\
\theta^4_{122}&=&\frac 1{4M^2}\biggl[4(\tau^2(S-2Q^2)-\tau(Q^2+S)+Q^2)F+4Q^2F_{z_1}
+\tau (8 m^2+(1+2\tau)Q^2)(F_{1+}-\tau(1+\tau)F_d)
\nonumber\\&&
+\tau^2 (\tau(2S-M^2)+2(1+\tau )X)F_{1+}-\tau^3(Q^2+\tau M^2)F_d+4\tau(\tau+1)^2F_{IR}\biggr],
\nonumber\\
\theta^5_{122}&=&-\frac{1+\tau}{4M^2} \biggl[\tau^2 (2F+(1+\tau)(\tau F_d+F_{1+}))-4F_{z_1}\biggr],   
\nonumber\\
\theta^3_{211}&=&2(Q^2+(\tau-1)S)F-2Q^2F_{z_1}+ \tau^2F_{IR}+\frac 2{1+\tau}(XF-Q^2F_{z_1}),
\nonumber\\
\theta^4_{211}&=&-\tau (2 F_{z_1}+\frac{\tau^2}2 (F_d \tau+F_{1+})),
\nonumber\\
\theta^2_{212}&=&\frac{\tau S_p}{4 M^2}\biggl[
(2Q^2(Q^2+4 m^2)+(\tau-1)(\lambda_S+\lambda_X))F_d-Q^2S_pF_{1+}
-\frac 1{1+\tau}(Q^2 S_pF_{1+}-2Q^2(Q^2+4 M^2)F
\nonumber\\&&
-(\lambda_S+\lambda_X)F_d)
\biggr],   
\nonumber\\
\theta^3_{212}&=&\frac1{4M^2}\biggl[ 
2(4\tau X(Q^2+M^2)+Q^2(\tau Q^2-4S_p))F+4Q^4F_{z_1}+(\tau^3((S_p+8m^2)M^2-3Q^4+8Q^2S-6S^2)
\nonumber\\&&
+2\tau (Q^2+4 m^2)(2\tau X-Q^2)+\tau S_p((1-\tau)M^2-Q^2))F_d+(8m^2Q^2
+2\tau\lambda_S
+\tau Q^2(6S-4Q^2-M^2)
\nonumber\\&&
-\tau ^2(Q^2 (M^2-Q^2)+4 X^2))F_{1+}
-\frac \tau{1+\tau}(Q^2 (M^2(8F_{z_1}-F_d)
+2Q^2(F+2F_{z_1}))
\nonumber\\&&
+(2M^2S-Q^2S_p)(4F+F_d)
+(\lambda _S+\lambda_X-M^2 Q^2)F_{1+} )
\biggr],
\nonumber\\
\theta^4_{212}&=&\frac 1{4M^2}
\biggl[2(\tau-1)(\tau S_p-2(\tau+2)Q^2)F+8(Q^2-\tau M^2)F_{z_1}+\tau^2(4(1+\tau)(\tau S-2m^2)
-\tau ^2M^2
\nonumber\\&&
-(1+5\tau+3\tau^2 )Q^2)F_d+\tau (\tau (2(1+2\tau)S-\tau M^2)-(3+5\tau+3\tau^2)Q^2+8 m^2)F_{1+}
\biggr],
\nonumber\\
\theta^5_{212}&=&\frac{1+\tau}{4 M^2} (4F_{z_1}-\tau^2(2 F+(1+\tau) (\tau F_d+F_{1+}))),
\nonumber\\
\theta^2_{221}&=&\frac {\tau S_p}{4M^2}\biggl[
Q^2S_pF_{1+}-(2Q^2(Q^2+4 m^2)+(\tau-1) (\lambda_S+\lambda_X))F_d
+\frac 1{1+\tau}(2 Q^2 (4 M^2+Q^2)F
\nonumber\\&&
+(\lambda_S+\lambda_X)F_d-Q^2S_p F_{1+} )\biggr],
\nonumber\\
\theta^3_{221}&=&\frac 1{4M^2}\biggl[
4Q^2((\tau M^2 -(1+2\tau)S_p)F+4Q^4F_{z_1} 
+\tau (2 (Q^2+4 m^2) (Q^2-2\tau X)
+2\tau^2(XS_p+\lambda_X)
\nonumber\\&&
+(1-\tau+\tau^2)M^2S_p)F_d 
+(2 \tau^2 (XS_p+\lambda_X)
+Q^2 ((1+\tau)(8m^2-\tau M^2)-2 Q^2-4 \tau S_p))F_{1+} \biggl]
\nonumber\\&&
+\frac \tau{4(1+\tau )}(Q^2 (4 F+8 F_{z_1}+F_{1+})-S_pF_d),
\nonumber\\
\theta^4_{221}&=&\frac 1{4M^2}\biggl[
2((1+\tau) (2+5 \tau)Q^2-2\tau (2+3\tau)S)F+4(1+2 \tau)Q^2F_{z_1}
+\tau^2(2(1+\tau) (Q^2-2\tau X+4m^2)
\nonumber\\&&
-\tau^2(S_p+M^2))F_d 
+\tau ((1+\tau)(8m^2-6\tau X)-\tau^2(Q^2+M^2)  )F_{1+} 
\biggr],
\nonumber\\
\theta^5_{221}&=&
\frac{\tau (\tau+1)}{4 M^2}(4(1+\tau)F+4 F_{z_1}+\tau (1+2\tau)(\tau F_d+F_{1+})),
\nonumber\\
\theta^3_{222}&=&\frac 1{4M^2}\biggl[Q^2(2(Q^2+2M^2)(4F_{z_1}-\tau F)
-S_p(\tau(1+\tau^2)F_d+4(3+\tau)F))
+(\tau(1+\tau) (\lambda_S+\lambda_X)
\nonumber\\&&
-(2+\tau)Q^2(Q^2-8 m^2)
)F_{1+} 
+\frac \tau{1+\tau}
\bigl(Q^2 (S_p(4 F+F_d)-2(Q^2+2M^2)(F+2F_{z_1}))-(\lambda_S+\lambda_X)F_{1+}\bigr)
\biggr],
\nonumber\\
\theta^4_{222}&=&\frac 1{4M^2}
\biggl[4(3+3\tau+\tau^2)Q^2F+4(3+2\tau)(Q^2F_{z_1}- \tau S F)
+\tau^2((1+\tau)Q^2-2 \tau^2 X)F_d
\nonumber\\&&
+\tau (2 (2+\tau) (4 m^2-\tau S)-(3-\tau-2\tau^2)Q^2)F_{1+}\biggr],
\nonumber\\
\theta^5_{222}&=&\frac{1+\tau}{4M^2}(2\tau(2+\tau)F+4(1+\tau)F_{z_1}+\tau^3 (\tau F_d+F_{1+})).
\end{eqnarray}
\hfill\rule[1ex]{1pt}{1.5ex}\rule[2.3ex]{\columnwidth}{1pt}
\end{strip}
Here $S_p=S+X=2S-Q^2$, and $F$ and $F_i$ ($i=d,1+,z_1,IR$) can be expressed through the invariant as
\begin{eqnarray}
F&=&\frac 1{\sqrt{\lambda_q }},
\nonumber\\[1mm]
F_d&=&\frac 1{2\pi\sqrt{\lambda_q }}\int\limits_0^{2\pi}\frac{d\phi_k}{z_1z_2}
=\frac 1\tau\Biggl(\frac 1{C_2}-\frac 1{C_1}\Biggr),
\nonumber\\[1mm]
F_{1+}&=&\frac 1{2\pi\sqrt{\lambda_q }}\int\limits_0^{2\pi}d\phi_k
\Biggl(\frac 1{z_1}+\frac 1{z_2}\Biggr)
=\frac 1{C_1}+\frac 1{C_2},
\nonumber\\[1mm]
F_{z_1}&=&\frac 1{2\pi\sqrt{\lambda_q }}\int\limits_0^{2\pi}z_1d\phi_k
\nonumber\\[1mm]
&=&\frac {Q^2(S_p-v)+\tau (S(Q^2+v)+2M^2Q^2)}{\lambda_q^{3/2}},
\nonumber\\[1mm]
F_{IR}&=&\frac{R^2}{2\pi\sqrt{\lambda_q}}\int\limits_0^{2\pi}d\phi_k{\cal F}_{IR}=\frac{\tau ^2S_p F_d-(2+\tau)Q^2F_{1+}}{2(1+\tau) },
\nn
\label{fff}
\end{eqnarray}
where ${\cal F}_{IR}$ is defined by Eq.~(\ref{fir}), while $z_{1,2}=kk_{1,2}/kp_1$ depends on $\phi_k$ as
\begin{eqnarray}
z_1&=&\frac1{\lambda_q}\Biggl[
Q^2(S_p-v)
+\tau ((Q^2+v)S+2M^2Q^2)
\nn&&
-2M\sqrt{\lambda_z}\cos\phi_k\Biggr],
\nn
z_2&=&\frac1{\lambda_q}\Biggl[
Q^2(S_p-v)
+\tau ((Q^2+v)(X-v)-2M^2Q^2)
\nn&&
-2M\sqrt{\lambda_z}\cos\phi_k\Biggr],
\nn
\lambda_z&=&(\tau^q_{max}-\tau)(\tau-\tau^q_{min})(Q^2(S(X-v)-M^2Q^2)
\nn&&
-m^2\lambda_q),
\end{eqnarray}
and $\tau^q_{max/min}$ defined by Eq.~(\ref{ta}).

At last
\begin{eqnarray}
C_1&=&[4m^2M^2(\tau^q_{max}-\tau)(\tau-\tau^q_{min})
\nonumber\\[1mm]
&&
+(Q^2+\tau S)^2]^{1/2},
\nonumber\\[1mm]
C_2&=&[4m^2M^2(\tau^q_{max}-\tau)(\tau-\tau^q_{min})
\nonumber\\[1mm]
&&
+(Q^2+\tau (v-X))^2]^{1/2}.
\label{c12}
\end{eqnarray}

\section{Calculation of $\delta_S$}
\label{ap2}
The real photon phase space in the dimensional regularization  has a form
\begin{eqnarray}
\frac{d^3k }{k_0}&\to& \frac{d^{n-1}k^\prime}{(2\pi \mu)^{n-4}k_0^\prime}
\nonumber\\
&=&
\frac{2\pi ^{n/2-1}k_0^{\prime n-3}dk_0^\prime \sin ^{n-3}\theta_k  d\theta_k }{(2\pi \mu)^{n-4}\Gamma(n/2-1)},
\label{dnk}
\end{eqnarray}
where $\theta_k$ is defined as the spatial angle between the photon three-momentum and 
${\bf k}_{1,2}^\prime $ or ${\bf k}_{s,x}^\prime $ that are introduced below, and
$\mu $ is an arbitrary parameter of the dimension of a mass. Here and later the upper
prime index means that the energy or three-momentum is defined in the system ${\bf p}_1{\bf + q=0}$.

The Feynman parametrization of the propagators in ${\cal F}_{IR}$:
\begin{eqnarray}
{\cal F}_{IR}
&=&\frac 1{4k^{\prime 2}_0}\int\limits_0^1dy \Biggl[\frac 2{1-x\beta_1}-\frac 2{1-x\beta_2}
-\frac S{k_{s0}^{\prime 2}(1-x\beta_s)^2}
\nonumber\\&&
    +\frac X{k_{x0}^{\prime 2}(1-x\beta_x)^2}\Biggr]
=\frac 1{4k^{\prime 2}_0}\int\limits_0^1dy {\mathcal F}(x,y),
\end{eqnarray}
where $y$ is the Feynman parameter, $x=\cos \theta_k $, $\beta _i=|{\bf k}^{\prime }_i|/k_{i0}^{\prime }$ for $i=1,2,s,x$,
$k_{s0}^{\prime }=yk_{10}^{\prime }+(1-y)p_{10}^{\prime }$ and
$k_{x0}^{\prime }=yk_{20}^{\prime }+(1-y)p_{10}^{\prime }$.

After substituting  it into definition of $\delta_S$ in (\ref{dsh}) and, using $\delta$ function,
integration of the obtained result  over the photon energy we found that
\begin{eqnarray}
\delta_S&=&\frac 1{2(4\mu \sqrt{\pi })^{n-4}\Gamma(n/2-1)}\int\limits_{-1}^1dx(1-x^2)^{n/2-2}
\nonumber\\&&\times
\int\limits_0^1dy {\mathcal F}(x,y)
\int\limits_0^{\bar{v}}\frac {dv}v\left(\frac vM\right)^{n-4}.
\end{eqnarray}

Then, 
the integration over $v$, and expansion of the obtained  expression into the Laurent series around
$n=4$ result in
\begin{eqnarray}
\delta_S&=&\delta_S^{IR}+\delta_S^1,
\end{eqnarray}
where
\begin{eqnarray}
\delta_S^{IR}&=&\frac 12\biggl[P_{IR}+\log \frac{\bar v}{\mu M }\biggr] 
\int\limits_0^1dy
\int\limits_{-1}^1dx
{\mathcal F}(x,y)
\end{eqnarray}
 and
\begin{eqnarray}
\delta_S^1&=&\frac 14
\int\limits_0^1dy
\int\limits_{-1}^1dx\log\biggl(\frac{1-x^2}4\biggr)
{\mathcal F}(x,y).
\end{eqnarray}
Here $P_{IR}$ is the infrared divergent term defined by Eq.~(\ref{pir}).
Since $k_{s0}^{\prime 2}-|{\bf k}^\prime_s|^2=m_y^2(S)$ and
$k_{x0}^{\prime 2}-|{\bf k}^\prime_x|^2=m_y^2(X)$
where
\begin{eqnarray}
m_y^2(S)=y(1-y)S+y^2m^2+(1-y)^2M^2,
\end{eqnarray}
the integration over $x$ and $y$ variables in $\delta_S^{IR}$ is performed explicitly:
\begin{eqnarray}
\delta_S^{IR}&=&2(XL_X-SL_S)\biggl[P_{IR}+\log \frac{\bar v}{\mu M }\biggr],
\end{eqnarray}
where the quantities $L_S$, and $L_X$ are defined by Eqs.~(\ref{lslx}).

For the covariant
analytical integration in $\delta_S^1$ we express
the initial and final lepton energies through the invariants:
\begin{eqnarray}
k_{10}^\prime=\frac{X }{2M},\qquad
k_{20}^\prime=\frac{S }{2M}.
\end{eqnarray}
As a result,
\begin{eqnarray}
\delta_S^1
&=&\frac 14 S\sqrt{\lambda_S}L_S^2
-\frac 14 X\sqrt{\lambda_X}L_X^2
\nonumber\\&&
+\frac S{\sqrt{\lambda_S}}{\rm Li}_2\biggl(\frac {2\sqrt{\lambda_S}}{S+\sqrt{\lambda_S}}\biggr)
-\frac X{\sqrt{\lambda_X}}{\rm Li}_2\biggl(\frac {2\sqrt{\lambda_X}}{X+\sqrt{\lambda_X}}\biggr)
\nonumber\\[2mm]&&
+S_\phi(k_2,p_1,p_2)
-S_\phi(k_1,p_1,p_2)
.
\end{eqnarray}

The function $S_\phi$ can be expressed through the  integration
over $x$ and $y$ as
\begin{eqnarray}
S_\phi(k_{1,2},p_1,p_2)&=&\frac {k_{1,2}p_1}2\int\limits_{-1}^1dx\int\limits_0^1dy
\frac{\log[(1-x^2)/4]}{k^{\prime 2}_{s0,x0}(1-x \beta_{s,x})^2}.
\nonumber\\
\end{eqnarray}
The explicit expression for $S_\phi$ can be found in Appendix B of \cite{lpcth}.

\end{document}